\pdfoutput=1
 \newcommand*{\ATLASLATEXPATH}{}
\documentclass[cernpreprint, texlive=2016, USenglish]{\ATLASLATEXPATH atlasdoc}

\usepackage{\ATLASLATEXPATH atlaspackage}
\usepackage{\ATLASLATEXPATH atlasbiblatex}

\usepackage{\ATLASLATEXPATH atlasphysics}

\usepackage{EXOT-2016-09-PAPER-defs}
\usepackage{comment}
\usepackage{caption}
\usepackage{multirow}

\DeclareCaptionStyle{capstyle}[justification=centering]{format=hang, labelfont={rm, bf}, textfont={rm} }
\mathchardef\mhyphen="2D

\AtlasTitle{Search for lepton-flavor violation in different-flavor, high-mass
final states in \boldmath{$pp$} collisions at
\boldmath{$\sqrt{s}=13$}~\si{\tera\electronvolt} with the ATLAS detector}
\usepackage{authblk}
\author{The ATLAS Collaboration}
\AtlasRefCode{EXOT-2016-09}
\PreprintIdNumber{CERN-EP-2018-137}
\AtlasAbstract{%
A search is performed for a heavy particle decaying into different-flavor,
dilepton pairs ($e\mu$, $e\tau$ or $\mu\tau$), using \SI{36.1}{\per\femto\barn}
of proton--proton collision data at $\sqrt{s}=13$~\si{\tera\electronvolt} 
collected in 2015--2016 by the ATLAS detector at the Large Hadron Collider. 
No excesses over the Standard Model predictions are observed. Bayesian lower
limits at the 95\% credibility level are placed on the mass of a \Zprime~boson,
the mass of a supersymmetric $\tau$-sneutrino, and on the threshold mass for
quantum black-hole production.
For the \Zprime~and sneutrino models, upper cross-section limits are converted
to upper limits on couplings, which are compared with similar limits from
low-energy experiments and which are more stringent for the $e\tau$ and
$\mu\tau$ modes.}
\hypersetup{pdftitle={ATLAS document},pdfauthor={The ATLAS Collaboration}}
 
\begin{document}

\maketitle

\clearpage

\section{Introduction}
\label{sec:introduction}
Direct charged-lepton flavor violation (LFV) is forbidden in the Standard Model
(SM) of particle physics but is allowed in many extensions of the SM. Many such
models predict new particles with LFV decays, such as \Zprime~bosons
\cite{bib:Langacker2008}, scalar neutrinos in $R$-parity-violating (RPV)
\cite{Farrar:1978xj,bib:RParitySUSY} supersymmetry (SUSY)
and quantum black holes (QBH) in low-scale gravity \cite{bib:Gingrich2009}.
Processes with flavor-violating dilepton decays are expected to produce pairs
of prompt, different-flavor leptons, a final state with a clear experimental
signature and a low background from SM processes.
The $\Z/\gamma^* \to \ell\ell$ process is an irreducible background for
same-flavor lepton searches but in different-flavor searches is limited to the
production and decay of a $\tau\tau$ pair. This paper describes a search for new
phenomena in final states with two leptons of different flavor using
\SI{36.1}{\per\femto\barn} of data from proton--proton ($pp$) collisions at
$\sqrt{s}=13$~\si{\TeV} at the Large Hadron Collider (LHC).

A common extension of the SM is the addition of an extra $U(1)$ gauge symmetry
resulting in a massive neutral vector boson known as a \Zprime~boson
\cite{bib:Langacker2008}. The search presented in this paper assumes a
\Zprime~boson with the same quark couplings and chiral structure as the SM $Z$
boson but only leptonic decays that violate Lepton Flavour Conservation are
allowed. The parameter $Q_{ij}$, where $i,j = 1 \ldots 3$ represent the three
lepton generations, gives the strength of the LFV couplings relative to the SM
$\ell\ell$ couplings. The ATLAS Collaboration placed lower limits of 3.3,
2.9, and \SI{2.7}{\TeV} on the mass of a \Zprime~boson decaying into $e\mu$,
$e\tau$, and $\mu\tau$ pairs, respectively, using \SI{3.2}{\per\femto\barn} of
the \SI{13}{\TeV} data \cite{EXOT-2015-20}, while the CMS Collaboration has
placed limits up to \SI{4.4}{\TeV} on a \Zprime~boson decaying into an $e\mu$
final state using \SI{35.9}{\per\femto\barn} \cite{CMS-EXO-16-058}. Following
the same methodology as in Ref.~\cite{EXOT-2015-20}, this paper assumes  only
one LFV coupling is different from zero at any time for the purpose of setting
limits on the cross-section times branching of each final state considered.
Polarization of $\tau$-leptons is not included in the model, but its impact on
the $\tau$-lepton acceptance is found to be negligible.

In RPV SUSY, the superpotential terms allowing LFV are expressed as
$\frac{1}{2}\lambda_{ijk}L_i L_j E^c_k + \lambda^{\prime}_{ijk}L_i Q_j D^c_k$,
where $L$ and $Q$ are the $SU(2)$ doublet superfields of leptons and quarks,
$E$ and $D$ are the $SU(2)$ singlet superfields of charged leptons and
down-type quarks, $\lambda$ and $\lambda^\prime$ are Yukawa couplings, and the
indices $i$, $j$, and $k$ denote generations. A $\tau$-sneutrino
($\tilde{\nu}_{\tau}$) may be produced in $pp$ collisions by $d\bar{d}$
annihilation and subsequently decay into $e\mu$, $e\tau$, or $\mu\tau$. Although
only $\tilde{\nu}_{\tau}$ is considered in this paper, results apply to any
sneutrino flavor. For the theoretical prediction of the cross-section times
branching ratio, the $\tilde{\nu}_{\tau}$ coupling to first-generation quarks
($\lambda^{\prime}_{311}$) is assumed to be 0.11 for all channels. As in the
\Zprime~model, each lepton-flavor-violating final state is considered
separately. It is assumed that
$\lambda_{312} = \lambda_{321} = 0.07$ for the $e\mu$ final state,
$\lambda_{313} = \lambda_{331} = 0.07$ for the $e\tau$ final state, and
$\lambda_{323} = \lambda_{332} = 0.07$ for the $\mu\tau$ final state.
These values are consistent with benchmark models used in previous ATLAS and
CMS searches \cite{EXOT-2015-20,SUSY-2013-06,CMS-EXO-13-002}. The CMS
Collaboration has recently excluded $R$-parity-violating supersymmetric models
below \SI{1.7}{\TeV} for
$\lambda_{132} = \lambda_{231} = \lambda^{\prime}_{311} = 0.01$
\cite{CMS-EXO-16-058}.

Various models introduce extra spatial dimensions to reduce the value of the
Planck mass and resolve the hierarchy problem. The search described in this
paper presents interpretations based on the Arkani-Hamed--Dimopoulos--Dvali
(ADD) model~\cite{bib:Hamed1998}, assuming $n=6$, where $n$ is the number of
extra dimensions, and on the Randall--Sundrum (RS) model \cite{bib:Randall1999}
with one extra dimension. Due to the increased strength of gravity at short
distances in these models, $pp$ collisions at the LHC could produce states
exceeding the threshold mass ($m_{\mathrm{th}}$) and form black holes. For the
models considered, $m_{\mathrm{th}}$ is assumed to be equivalent to the
extra-dimensional Planck scale. For masses beyond $3$--$5 m_{\mathrm{th}}$, it is
expected that thermal black holes would be produced
\cite{bib:Dimopoulos2001,bib:Giddings2001},
characterized by high-multiplicity final states. The search presented in this
paper focuses on the mass region below 3--5 $m_{\mathrm{th}}$, known as the quantum
gravity regime \cite{bib:MeadeRandall2007,bib:Gingrich2008,bib:Calmet2008},
where production of non-thermal (or quantum) black holes is expected and these
black holes could decay into two-particle final states, producing the topology
investigated in this paper.
Non-thermal quantum black holes would have a continuum mass distribution from
$m_{\mathrm{th}}$ up to the beginning of the thermal regime. For the models
considered in this paper, the thermal regime is assumed to start at
$3 m_{\mathrm{th}}$. The decay of quantum black holes would be governed by a yet
unknown theory of quantum gravity. The two main assumptions of the
extra-dimensions models considered \cite{bib:Gingrich2009} in this paper are
(a) gravity couples with equal strength to all SM particle degrees of freedom
and (b) gravity conserves local symmetries (color and electric charge) but can
violate global symmetries such as lepton-flavor and baryon-number conservation.
Following these assumptions, the branching ratio to each final state
is calculated. QBHs decaying into different-flavor, opposite-charge lepton pairs
are created via $q\bar{q}$ or $gg$ annihilation.
The branching ratio to $\ell\ell^\prime$ is 0.87\% (0.34\%) for a $q\bar{q}$
($gg$) initial state \cite{bib:Gingrich2009}. These models were used in
previous ATLAS and CMS searches for quantum black holes in dijet
\cite{EXOT-2013-11,EXOT-2015-02,CMS-EXO-12-059}, lepton+jet
\cite{EXOT-2013-03}, photon+jet \cite{EXOT-2012-22}, $e\mu$
\cite{CMS-EXO-16-058}, and same-flavor dilepton \cite{EXOT-2012-23} final
states.

\section{The ATLAS detector}
\label{sec:ATLASDetector}
The ATLAS detector \cite{PERF-2007-01} is a general-purpose particle
detector with approximately forward--backward symmetric cylindrical
geometry.\footnote{ATLAS uses a right-handed coordinate system with its origin
  at the nominal interaction point (IP) in the center of the detector and the
  $z$-axis along the beam pipe. The $x$-axis points from the IP to the center
  of the LHC ring, and the $y$-axis points upward. The $x$--$y$ plane is
  referred to as the transverse plane, used to define quantities such as the
  transverse momentum ($p_{\mathrm{T}}$). Cylindrical coordinates $(r,\phi)$ are
  used in the transverse plane, $\phi$ being the azimuthal angle around the
  $z$-axis. The pseudorapidity is defined in terms of the polar angle $\theta$
  as $\eta=-\ln\tan(\theta/2)$. Angular distance is measured as
  $\Delta R = \sqrt{ (\Delta\eta)^{2} + (\Delta\phi)^{2} }$.} It is composed
of four main components, each responsible for identifying and reconstructing
different types of particles: the inner detector (ID), the electromagnetic
and hadronic calorimeters, and the muon spectrometer (MS).
Each of the subdetectors is divided into two components, barrel and endcap, 
to provide coverage close to $4\pi$ in solid angle. In addition, two magnet
systems allow charge and momentum measurements: an axial magnetic field of
\SI{2.0}{\tesla} provided by a solenoid surrounding the ID and a toroidal
magnetic field  for the MS. The ID, the closest component to the interaction
point, is used to reconstruct the trajectories of charged particles in the
region $|\eta|<2.5$ and measure their momenta. It is composed of three
subsystems: (1) the silicon pixel detector, including an additional inner layer
at a radius of 3.2~\si{\centi\meter} added in 2015
\cite{bib:IBL_TDR,ATLAS:1451888}, (2) the semiconductor tracker, used in
conjunction with the silicon pixel detector to determine primary and secondary
vertices with high precision thanks to their high granularity, (3) the
transition radiation tracker, providing additional tracking in the region
$|\eta|<2.0$ and electron identification.

The calorimeter system covers the pseudorapidity range $|\eta| < 4.9$.
Within the region $|\eta|< 3.2$, electromagnetic calorimetry is provided by
barrel and  endcap high-granularity lead/liquid-argon (LAr) electromagnetic
calorimeters, with an additional thin LAr presampler covering $|\eta| < 1.8$,
to correct for energy loss in material upstream of the calorimeters.
Hadronic calorimetry is provided by the steel/scintillating-tile calorimeter,
segmented into three barrel structures within $|\eta| < 1.7$, and two
copper/LAr hadronic endcap calorimeters. The solid angle coverage is completed
with forward copper/LAr and tungsten/LAr calorimeter modules optimised for
electromagnetic and hadronic measurements respectively.

Surrounding the calorimeter system, the MS is the subdetector furthest from the
interaction point. It consists of three layers of precision tracking chambers
and fast detectors for triggering on muons. Tracking coverage is provided for
$|\eta|<2.7$ by three layers of precision drift tube chambers, with cathode
strip chambers in the innermost layer for $|\eta|>2.0$, while trigger coverage
is provided by resistive plate and thin gap chambers for $|\eta|<2.4$.

The trigger and data-acquisition system is based on two levels of online event
selection \cite{bib:trigger2015}: the level-1 trigger and the high-level
trigger. The level-1 trigger is hardware-based and uses a subset of detector
information to provide quick trigger decisions and reduce the accepted rate to
\SI{100}{\kilo\hertz}. The high-level trigger is software-based and exploits
the full detector information to further reduce the acceptance rate to about one
\si{\kilo\hertz}.

\section{Data and simulated samples}
\label{sec:dataMC}
The data sample used for this analysis was collected during 2015 and
2016 from $pp$ collisions at a center-of-mass energy of \SI{13}{\TeV}. After
selecting periods with stable beams and applying data-quality requirements, the
total integrated luminosity is \SI{36.1}{\per\femto\barn} with an uncertainty
of 2.1\%, derived following a methodology similar to that detailed in
Ref.~\cite{bib:LumiNew} from a calibration of the luminosity scale using
$x$--$y$ beam-separation scans.

The $pp\to\Zprime\to\ell\ell^{\prime}$ samples were generated at leading
order (LO) using the generator \Pythia~8.186~\cite{bib:Pythia8}
with the NNPDF23LO \cite{bib:Ball2012} parton distribution function (PDF) set
and the A14~\cite{ATL-PHYS-PUB-2014-021} set of tuned parameters. Signal samples
for 25 mass points ranging from \SI{0.5}{\TeV} to \SI{5}{\TeV} were generated
in \SI{0.1}{\TeV} steps from 0.5 to \SI{2}{\TeV}, \SI{0.2}{\TeV} steps from
2 to \SI{3}{\TeV}, and \SI{0.5}{\TeV} steps from 3 to \SI{5}{\TeV}. The
production cross-section was calculated with the same generator used for
simulation. The cross-section and signal shape in the dilepton invariant mass
distribution were corrected from LO to next-to-next-to-leading order (NNLO) in
the strong coupling constant with a rescaling that depends on the dilepton
invariant mass and which was computed with VRAP 0.9~\cite{Anastasiou:2003ds}
and the CT14NNLO PDF \cite{bib:Dulat2015} set. This correction is applied as a
multiplicative factor of about 0.98 at a dilepton invariant mass
$m_{\ell\ell^{\prime}}$ of \SI{3}{\TeV}. No mixing of the $Z^\prime$ boson with the
$Z$ and $\gamma^*$ bosons is included.

The $d\bar{d}\to\tilde{\nu}_{\tau}\to\ell\ell^{\prime}$ samples were
generated at LO with \Madgraph~v2.3.3~\cite{bib:MadGraph5}
interfaced to the \Pythia~8.186 parton shower model with the NNPDF23LO PDF set
and the A14 tune. The signal samples were generated at the
same masses as for the \Zprime~model described above. The cross-section was
calculated at LO with the same generator used for simulation and corrected
to next-to-leading order (NLO) using \LoopTools~v2.2~\cite{bib:LoopTools}.

The $pp\to$ QBH  $\to\ell\ell^{\prime}$ samples were generated with the program
\QBH~3.00 \cite{bib:QBHManual}  using the CTEQ6L1 \cite{bib:CTEQ6L1} PDF set
and the A14 tune, for which \Pythia~8.183 provides showering and
hadronization. For each extra-dimensional model, eleven $m_{\mathrm{th}}$ points
in \SI{0.5}{\TeV} steps were produced: from 3 to \SI{8}{\TeV} for the ADD $n=6$
model and from 1 to \SI{6}{\TeV} for the RS $n=1$ model. The production
cross-section was calculated with the same generator used for simulation.
These two models differ in the number and nature of the additional extra
dimensions (large extra dimensions for ADD and one highly warped extra
dimension for RS). In particular, the ADD model produces black
holes with a larger gravitational radius and hence the parton--parton
cross-section for this model is larger than for the RS model. Therefore, the
$m_{\mathrm{th}}$ range of the generated samples differs for the two models.

The SM background in the LFV dilepton search is due to several processes
which produce a final state with two different-flavor leptons. For the $e\mu$
mode, the dominant background contributions originate from \ttbar and
single-top production, with the subsequent decays of the top quark producing
leptonically decaying \W~bosons. Other backgrounds originate from diboson
(\WW, \WZ, and \ZZ) production, and $\tau$-lepton pair production
$q\bar{q} \to\Z/\gamma^{*}\to\tau\tau$, which both produce different-flavor
final states, through the leptonic decays of the \W~and \Z~bosons or the
$\tau$-leptons. They contribute about 15\% and 1\% of the background,
respectively. Multijet and \W+jets processes contribute due to the
misidentification of jets as leptons and are the dominant background for the
final states with a $\tau$-lepton.

Backgrounds from top-quark production include \ttbar and single-top with an
associated \W~boson ($t\W$). Both were generated at NLO using the
\powhegbox~\cite{bib:Nason2004,bib:Frixione2007,bib:Alioli2010xd}
generator (v2 for \ttbar and v1 for single-top) with the CT10 \cite{bib:Lai2010}
PDF set used in the matrix-element calculations. \Pythia~6.4.28
\cite{bib:Sjostrand2006za} and the corresponding Perugia 2012 tune
\cite{bib:Skands2010ak} were used to simulate the parton shower,
hadronization, and the underlying event. Top quarks were decayed using
\madspin~\cite{bib:Artoisenet2012st}, preserving all spin correlations. The
\textit{h$_{\mathrm{damp}}$} parameter, which controls the $p_{\mathrm{T}}$ of the
first emission beyond the Born configuration in \powhegbox, was set to the mass
of the top quark. The main effect of this parameter is to regulate the
high-$p_{\mathrm{T}}$ emission against which the \ttbar system recoils. The mass
was set to the top quark mass of \SI{172.5}{\GeV}. The EvtGen 1.2.0
program~\cite{bib:EvtGen} was used for the properties of $b$- and $c$-hadron
decays. A value of 831$^{+20}_{-29}$ (scale) $^{+35}_{-35}$
(PDF+$\alpha_{\mathrm{S}}$) $^{+23}_{-22}$ (mass uncertainty) \si{\pico\barn} is
used for the \ttbar production cross-section, computed with
\toppp2~\cite{bib:Czakon2011}, incorporating NNLO QCD corrections, including
resummation of next-to-next-to-leading logarithmic (NNLL) soft gluon terms. A
$\W t$ production cross-section of $71.7\pm3.8$~\si{\pico\barn} is used, as
computed in Ref.~\cite{bib:SingleTopNNLO} to approximately NNLO (NNLL+NLO)
accuracy.

Diboson processes producing at least two charged leptons were simulated
using the \Sherpa~2.2.2 generator~\cite{bib:Gleisberg2008ta}. The matrix
elements contain all diagrams with four electroweak vertices. Fully leptonic
decays were calculated for up to one parton (four leptons, or two leptons and
two neutrinos) or zero partons (three leptons and one neutrino) at NLO and up
to three partons at LO using the \comix~\cite{bib:Gleisberg2008fv} and
\openloops~\cite{bib:Cascioli2011va} matrix-element generators and merged with
the \Sherpa~parton shower \cite{bib:Schumann2007} using the ME+PS@NLO
prescription \cite{Hoeche:2012yf}. The CT10 PDF set was used
in conjunction with the default parton shower tuning provided by the authors of
\Sherpa. Inclusive cross-section values of 1.28, 4.51, and 10.64~\si{\pico\barn}
are used for $ZZ$, $WZ$, and $WW$ production, respectively.

Events containing \W~or \Z~bosons are generated using \powhegbox~v2 interfaced
to the \Pythia~8.186 parton shower model. The CT10 PDF set is used in the matrix
element. The AZNLO set of tuned parameters \cite{bib:AZNLO} is used, with PDF
set CTEQ6L1, for the modelling of non-perturbative effects. The EvtGen 1.2.0
program  is used for the properties of b- and c-hadron decays. Photos++
3.52~\cite{Davidson:2010ew} is used for QED emissions from electroweak vertices
and charged leptons. The \W/\Z~samples are normalised with the NNLO cross
sections. This background contribution is normalized to an inclusive
cross-section of \SI{1.9}{\nano\barn}, calculated for
$m_{\ell\ell} >$\SI{60}{\GeV}.

Processes such as \W+jets and multijet production with jets that are
misidentified as leptons were estimated through a combination of data-driven
methods and simulation, detailed in Section~\ref{sec:background}. The \W+jets
contribution was estimated with the aid of the \Sherpa~2.2.2 simulated samples.
Matrix elements were calculated for up to two partons at NLO and four partons
at LO using Comix and OpenLoops and merged with the \Sherpa~parton
shower~\cite{bib:Schumann2007} according to the ME+PS@NLO
prescription~\cite{Hoeche:2012yf}.
The overall cross-section times branching ratio for the
$W^{\pm}\to\ell^{\pm}\nu$+jets events is taken to be 59.6~\si{\nano\barn}.

For all samples used in this analysis, the effects of multiple proton--proton
interactions per bunch crossing (pileup) were included by overlaying
minimum-bias events simulated with \pythia8.186 using the ATLAS A14 set of
tuned parameters~\cite{bib:A14} and reweighting the simulated
events to reproduce the distribution of the number of interactions per bunch
crossing observed in the data. The generated events were processed with the
ATLAS simulation infrastructure~\cite{SOFT-2010-01}, based on
\geant4~\cite{bib:Geant4}, and passed through the trigger simulation and the
same reconstruction software used for the data.

\section{Event reconstruction and selection}
\label{sec:obj_evt_selection}
This search is optimized  to look for new phenomena in the high mass range.
Events are selected if they satisfy a single-muon or single-electron trigger
with a $p_{\mathrm{T}}$ threshold of \SI{50}{\GeV} for muons and 60 or
\SI{120}{\GeV} for electrons. The single-electron trigger with higher
$p_{\mathrm{T}}$ threshold has a looser identification requirement, resulting
in an increased trigger efficiency at high $p_{\mathrm{T}}$.

Electron candidates are formed by associating the energy in clusters of cells
in the electromagnetic calorimeter with a track in the ID \cite{PERF-2016-01}.
A likelihood discriminant suppresses contributions from hadronic jets,
photon conversions, Dalitz decays, and semileptonic heavy-flavor hadron.
The likelihood discriminant utilizes lateral and longitudinal
calorimeter shower shapes plus tracking and cluster--track matching quantities.
The discriminant criterion is a function of the \pT\ and $|\eta|$ of
the electron candidate. Two operating points are used in this analysis, as
defined in Ref.~\cite{ATLAS-CONF-2016-024}: \emph{Medium} and \emph{Tight}.
The \emph{Tight} working point (85\% efficient at $\pT = 65$~\si{\GeV}
determined with $\Z \to ee$ events)
is required for electron candidates. Electron candidates must have
$\pT > 65$~\si{\GeV} and $|\eta|<2.47$, excluding the region
$1.37<|\eta|<1.52$, where the energy reconstruction performance is degraded due
to the presence of extra inactive material. Further requirements are made on the
transverse and longitudinal impact parameters of the track, which is the
distance between the $z$-position of the point of closest approach of the track
in the ID to the beamline and the $z$-coordinate of the primary vertex relative
to the primary vertex of the event ($d_0$ and $\Delta z_0$). The requirements
are the following: $|d_{\mathrm{0}}/\sigma_{d_{\mathrm{0}}}|<5$ and
$|\Delta z_{\mathrm{0}}\sin\theta|<0.5$~\si{\milli\meter}.
Candidates are required to satisfy relative track-based (as defined above for
muon candidates) and calorimeter-based isolation requirements with an efficiency
of 99\% to suppress background from non-prompt electrons originating from
heavy-flavor semileptonic decays, charged hadrons, and photon conversions from
$\pi^{\mathrm{0}}$ decays. The sum of the calorimeter transverse energy deposits
(excluding the electron itself) in an isolation cone of size $\Delta R=0.2$
divided by the electron $p_{\mathrm{T}}$ is the discriminant used in the
calorimeter-based isolation criterion. For the reducible background
estimation, electron candidates passing the \emph{Medium} working point (95\%
efficient at $p_{\mathrm{T}}=65$~\si{\GeV} determined with $\Z/\gamma^* \to ee$
events) are referred to as ``loose electrons''.

Candidate muon tracks are initially reconstructed independently in the ID and
the MS. The two tracks are input to a combined fit which takes
into account the energy loss in the calorimeter and multiple scattering.
Muon identification is based on information from both the ID and MS to ensure
that muons are reconstructed with the optimal momentum resolution up to very
high $p_{\mathrm{T}}$ using the \emph{High}-$p_{\mathrm{T}}$ operating
point~\cite{PERF-2015-10}. Only tracks with hits in each of the three stations
of the muon spectrometer are considered. This provides a muon transverse
momentum resolution of about 10\% at 1~\TeV. Moreover, muon candidates are
required to be within the ID acceptance region\footnote{For the $\mu\tau$
channel, the muon acceptance is limited by the coverage of the muon trigger
system ($|\eta| < 2.4$).} of $|\eta| <$ 2.5, fulfill
$|d_{\mathrm{0}}/\sigma_{d_{\mathrm{0}}}|<3$ and
$|\Delta z_{\mathrm{0}}\sin\theta|<0.5$~\si{\milli\meter},
have a transverse  momentum larger than $65$~\si{\GeV}, and fulfill a
track-based isolation criterion with an efficiency of 99\% over the full range
of muon momenta to further reduce contamination from non-prompt muons.
The scalar sum of the transverse momenta of tracks (excluding the muon itself)
in an isolation cone of size $\Delta R=0.2$ divided by the muon $p_{\mathrm{T}}$
is the discriminant used in the track-based isolation criterion. For the
reducible background estimation, muon candidates fulfilling all selection
criteria except the isolation criterion are called ``loose muons''.

Jets are reconstructed using the anti-$k_{t}$ algorithm \cite{bib:Cacciari2008}
with a radius parameter of 0.4 using energy clusters
\cite{PERF-2014-07} of calorimeter cells as input. Jet
calibrations~\cite{PERF-2016-04} derived from $\sqrt{s}=13$~\si{\TeV}
simulated data and from collision data taken at 13~\si{\TeV} are used to correct the jet energies and directions to those of the particles from the hard-scatter interaction.

Hadronic decays of $\tau$-leptons are composed of a neutrino and a set of
visible decay products ($\tau_{\text{had-vis}}$), typically one or three charged
pions and up to two neutral pions. The reconstruction of $\tau$-leptons and
their visible hadronic decay products starts with jets reconstructed from
topological clusters~\cite{bib:TauReconstructionRun2}.
The $\tau_\mathrm{had \mhyphen vis}$ candidates must have $|\eta|<2.5$ with the
transition region between the barrel and endcap calorimeters
($1.37<|\eta|<1.52$) excluded, a transverse momentum greater than
\SI{65}{\GeV}, one or three associated tracks with $\pm$1 total electric charge.
Their identification is performed using a multivariate algorithm that employs
boosted decision trees (BDT) using shower shape and tracking information to
discriminate against jets. All  $\tau_\mathrm{had \mhyphen vis}$  candidates are
required to fulfill the ``loose'' identification requirements of
Ref.~\cite{ATLAS-CONF-2017-029}. An additional dedicated likelihood-based veto
is used to reduce the number of electrons misidentified as $\tau_{\text{had-vis}}$
candidates.  For the purpose of the reducible background estimation,
$\tau_\mathrm{had \mhyphen vis}$ candidates failing the ``loose'' identification
requirements are used.

Jets containing $b$-hadrons ($b$-jets) are identified with a $b$-tagging
algorithm based on a multivariate technique \cite{ATL-PHYS-PUB-2015-022}.
Operating points are defined by a single value in the domain of discriminant
outputs and are chosen to provide a specific $b$-jet efficiency in an
inclusive \ttbar sample. The employed working point has an efficiency of 77\%
and rejection factors of 6 and 134 for charm and light-quark/gluon jets,
respectively \cite{ATL-PHYS-PUB-2015-022}.

Only events with exactly two different-flavour leptons are chosen. As such,
there is no overlap between the three channels considered: $e\mu, e\tau,
\mu\tau$.  They must have a reconstructed primary vertex, defined as the vertex
whose constituent tracks have the highest sum of $p_{\mathrm{T}}^2$, and exactly
two reconstructed different-flavor lepton candidates meeting the above-mentioned
criteria. Events with an additional electron, muon or
$\tau_\mathrm{had \mhyphen vis}$ are vetoed. For the $e\mu$ channel only,
events with an extra electron or muon fulfilling the ``loose'' criteria are
also vetoed, including events used for the purpose of the reducible background
estimation.
For all three channels, the lepton candidates must be back-to-back in
the transverse plane with $\Delta\phi(\ell,\ell^{\prime})>2.7$. The invariant
mass of the dilepton pair is used as the discriminant. No requirement
is made on the respective charges of the leptons since it reduces the signal
efficiency by as much as 6\% for the highest-mass signals considered due to
charge misassignment without a significant effect on the background rejection.
To account for differences between data and simulation, corrections are applied
to the lepton trigger, reconstruction, identification, and isolation
efficiencies as well as the lepton energy/momentum resolution and scale
\cite{ATLAS-CONF-2016-024,PERF-2016-01,PERF-2015-10,ATLAS-CONF-2017-029}.

Double-counting of leptons and jets is avoided by applying an overlap removal
algorithm based on the $\Delta R $ distance metric.  First, jets within
$\Delta R < 0.2$ of any identified muons or electrons are removed. Then, any
muons and electrons within $0.2 < \Delta R < 0.4$ from the jet axis are removed.

The missing transverse momentum vector ($\vec{E}_{\mathrm{T}}^{\mathrm{miss}}$) is
defined as the negative vector sum of the transverse momenta of all identified
physics objects and an additional soft term. The soft term is constructed
from all tracks that are associated with the primary vertex, but not with any
selected physics object. In this way, the missing transverse momentum
incorporates the best calibration of the jets and the other identified physics
objects, while maintaining pileup independence in the soft
term~\cite{bib:METperfo}.

The dominant background for the $e\mu$ channel is \ttbar production, which can
be suppressed by rejecting events that contain one or more $b$-jets ($b$-veto).

For a \Zprime~boson with a mass of \SI{1.5}{\TeV}, the fractions of events that
pass all of the selection requirements are approximately 45\%, 45\%, 20\%, and
15\% for the $e\mu$, $e\mu$ with $b$-veto, $e\tau$, and $\mu\tau$ final states,
respectively.

For the reducible background estimation in the $e\tau$ and $\mu\tau$ channels,
the transverse mass $m_{\mathrm{T}}$ of a lepton and the missing transverse
momentum is defined as

\begin{equation*}
  \label{eq:mT}
  m_{\mathrm{T}} = \sqrt{2 p_{\mathrm{T}}E_{\mathrm{T}}^{\mathrm{miss}}
                 [1-\cos\Delta\phi(\ell,E_{\mathrm{T}}^{\mathrm{miss}})]}\;,
\end{equation*}
where $p_{\mathrm{T}}$ is the transverse momentum of the lepton,
$E_\mathrm{T}^{\mathrm{miss}}$ is the magnitude of the missing transverse momentum
vector, and $\Delta\phi(\ell,E_{\mathrm{T}}^{\mathrm{miss}})$ is the azimuthal angle
between the lepton and $\vec{E}_{\mathrm{T}}^{\mathrm{miss}}$ directions.

For the dilepton mass calculation in the $e\tau$ and $\mu\tau$ channels, the
missing momentum from the neutrino in the hadronic decay of a  $\tau$-lepton is
estimated and added to the four-momentum of the $\tau_{\mathrm{had \mhyphen vis}}$
candidate. At the considered momenta, the hadronic decay of the $\tau$-lepton
results in the neutrino and the resultant jet being nearly collinear. The
neutrino four-momentum is reconstructed from the magnitude of the missing
transverse momentum and the direction of the $\tau_{\mathrm{had \mhyphen vis}}$
candidate. This technique significantly improves the dilepton mass resolution
and search sensitivity \cite{SUSY-2013-06}. For a simulated \Zprime~boson with
a mass of \SI{2}{\TeV}, the mass resolution improves from  8\% (17\%) to 4\%
(12\%) in the $e\tau$ ($\mu\tau$) channel.

\section{Background estimation}
\label{sec:background}
The background processes for this search can be divided into two categories:
reducible and irreducible. The latter is composed of processes which
produce two different-flavor prompt leptons, including
$\Z/\gamma^* \to\tau\tau$, \ttbar, single-top, and diboson production. These
processes are modeled using simulated samples and normalized to their
theoretically predicted cross-sections. Reducible backgrounds originate from
jets mis-reconstructed as leptons and require the use of data-driven
techniques. The contribution from reducible backgrounds is small in the $e\mu$
channel, about 5\%, whereas in the $e\tau$ and $\mu\tau$ channels they are
the leading components and make up 50$-$60\% of the total background.

\subsection{Top-quark background extrapolation}
\label{subsec:topquark}

The simulated samples used to estimate single-top-quark and \ttbar production
are statistically limited for dilepton invariant masses above \SI{1}{\TeV}.
Therefore, for $m_{\ell\ell'}>$700~\si{\GeV}, the \ttbar plus single-top
contributions are evaluated using monotonically decreasing functions
fitted to the $m_{\ell\ell'}$ distribution. Two functional forms are chosen
for their stability when varying the fit range and for the quality of the fit:

\begin{equation*}
\label{eq:djet-monimial}
  a \cdot m_{\ell\ell^{\prime}}^b \cdot
  m_{\ell\ell^{\prime}}^{c\cdot\mathrm{ln}(m_{\ell\ell^{\prime}})}
  \quad \mathrm{and} \quad \frac{a}{(m_{\ell\ell^{\prime}}+b)^c}\;,
\end{equation*}
where $a$, $b$ and $c$ are free parameters in the fit. To account for fit 
variations, the lower and upper limits of the fit range were varied
in 25~\GeV\ steps between 200$-$300~\GeV\ and 1000$-$1200 (800$-$1000)~\GeV\
for $e\mu$ ($e\tau$ and $\mu\tau$).
The nominal extrapolation is taken as the average of all the tested fit ranges
using both functional forms. The extrapolation is found to agree within
statistical uncertainties with the simulated data.
For each mass bin, the up and down variations obtained from varying the fit
parameters are combined in quadrature with the uncertainty of the fit range
variation. This uncertainty is 32\% at 2~\si{\TeV} for the $e\mu$ channel.

\subsection{Reducible background}
\label{subsec:RecBackground}

The main reducible backgrounds are \W+jets and multi-jet production. The
contribution to the reducible background from muons originating from decays of
hadrons in jets is found to be negligible compared with the contribution from
fake electrons and $\tau$-leptons. Therefore, in the $e\mu$ channel, where
reducible backgrounds are a small contribution to the total, non-prompt muons 
are neglected and only events with one prompt muon and a jet faking an electron
are considered. In channels involving taus, however, reducible backgrounds are
more significant, and so contributions from both electrons and muons, primarily
non-prompt leptons from heavy flavor decays, are taken into account. However,
the dominant source of reducible background in these channels remains fake taus
from gluon-initiated jets.

\subsubsection{$e\mu$ channel: matrix method}
\label{sec:emu_matrix}

For the $e\mu$ channel, the matrix method is employed, as detailed in
Ref.~\cite{EXOT-2012-23}. The selection criteria are loosened for electron
candidates to create a sample of events with a muon and a ``loose'' electron
as defined in Section~\ref{sec:obj_evt_selection}. These events are referred to
as ``loose'', while those in which both the electron and muon pass all
selections are ``tight''. The probability of a ``loose'' electron matched to a
generated electron to pass the full object selection (the ``real efficiency'')
is evaluated from $Z\to ee$ simulated events, while  the probability that a jet
is misidentified as an electron (the ``fake rate'') is obtained in a
multijet-enriched data sample. To suppress the~\W+jets
contribution to the multijet-enriched sample, its events are required to have
$E_{\mathrm{T}}^{\mathrm{miss}}<25$~\si{\GeV} and $m_{\mathrm{T}}<50$~\si{\GeV} as well
as to pass the signal region selection outlined in
Section~\ref{sec:obj_evt_selection}.
Both the real efficiency and the fake rate are determined as a
function of $p_{\mathrm{T}}$, and they are used to estimate the reducible
background contribution by solving a system of linear equations involving the
numbers of loose and tight events in the signal region.
Residual contaminations from W+jets and other SM background processes (top,
diboson, and $Z\to\ell\ell$) in the multi-jet CR are subtracted using
simulation. This background is estimated up to around \SI{1.5}{\TeV}, where
there are no data. However, at this stage, the expectation of this background
is well below one event, and generally negligible compared to diboson and top
quark processes.

The uncertainties associated with the matrix method are evaluated by considering
systematic effects on the electron fake rate. Uncertainties of the real electron
efficiency have a negligible impact on the estimation and are not considered.
The systematic uncertainties in the fake rate include
\begin{itemize}
\item{the choice of multijet-enriched region,}
\item{uncertainties on the MC subtraction in the multijet-enriched region, and}
\item{the difference in the fake rates obtained using this method and those
obtained from simulated \W+jets events.}
\end{itemize}
The overall uncertainty of the $e\mu$ reducible background is about 30\%.
Given that in the $e\mu$ channel this contribution is about 7\% of the total
background over the invariant mass range considered, the uncertainties in the
estimation method have a small impact on the results.

\subsubsection{$e\tau$ and $\mu\tau$ channels: \W+jets estimate}
\label{sec:Wjets_bkg}

The dominant background for the $e\tau$ and $\mu\tau$ channels is the \W+jets
process, where a jet is misidentified as a $\tau_\mathrm{had \mhyphen vis}$
candidate. It is estimated using simulated events with each jet weighted by
its probability to pass the $\tau$-lepton identification as measured in data.
This not only ensures the correct fake rate but also improves the statistical
precision of the estimate, since events failing the $\tau$-lepton identification
requirements are not discarded. The $\tau_\mathrm{had \mhyphen vis}$ fake rate
is measured in a \W$\to e$/$\mu$+jets control region as a function of the
$p_{\mathrm{T}}$, $\eta$, and number of tracks of the
$\tau_\mathrm{had \mhyphen vis}$ candidates.
The~\W+jets-enriched control region uses the same selection as the signal
region, but reverses the back-to-back criterion to
$\Delta\phi(\ell,\ell^{\prime}) < 2.7$, and uses $\tau$-leptons fulfilling all
requirements except identification, although a minimum requirement on the BDT
discriminant is retained. Only events with exactly one electron or muon
fulfilling all selection criteria, as well as $m_{\mathrm{T}}>80$~\si{\GeV} to
enrich the \W+jets contribution, are used. Events where the invariant mass of
the $e$ or $\mu$ and the $\tau_\mathrm{had \mhyphen vis}$ candidate is between
80 and \SI{110}{\GeV} are vetoed to reduce contamination from $Z$ boson decays.
Contributions from non-\W+jets processes are subtracted using simulation.
The $\tau_\mathrm{had \mhyphen vis}$ candidates present in the remaining events
are dominated by jets. The contribution of events with non-prompt electrons is
estimated from simulation and found to be less than 1\%.
The $\tau_\mathrm{had \mhyphen vis}$  fake rate is defined as the fraction of
$\tau_\mathrm{had \mhyphen vis}$  candidates in the sample  that also pass the
$\tau_\mathrm{had \mhyphen vis}$  identification. This  rate is used to weight
simulated \W+jets events. The resulting distribution obtained for the \W+jets
is validated in the~\W+jets-enriched control region, where good agreement is
found between data and the expected SM background processes.

The uncertainties in the $\tau_\mathrm{had \mhyphen vis}$ fake rate are
evaluated from
\begin{itemize}
\item{the modeling of the ``loose'' $\tau_\mathrm{had \mhyphen vis}$
identification requirement in simulation,}
\item{the statistical uncertainty of the data-driven estimation of the
$\tau$-lepton fake rate, and}
\item{the differences in $\tau$-lepton fake rate between signal and control
regions.}
\end{itemize}

These errors are detailed in the following paragraphs.

The $\tau_\mathrm{had \mhyphen vis}$ fake rate is re-evaluated when removing the
$m_{\mathrm{T}}>80$~\si{\GeV} requirement to check the contamination from
non-\W+jets processes. The effect on the fake rate and the final estimation of
the \W+jets background is about 2\%.

The statistical uncertainty of the fake rate in the control regions is
propagated through the estimate. The impact is small at low $m_{\ell\tau}$ but is
the leading uncertainty of the fake rate in the range \
$m_{\ell\tau} > 1$~\si{\TeV}.

The jet composition of the fake $\tau_\mathrm{had \mhyphen vis}$ background is
evaluated from simulated \W+jets events. The control region where the
$\tau_\mathrm{had \mhyphen vis}$ fake rate is evaluated should have a a jet
composition similar to that in the signal region. Therefore, \W+jets simulated
events are used to investigate the difference between the fake rates measured
in the \W+jets control and signal regions. The comparison reveals a slightly
lower fake rate in the signal region, consistent with the lower expected gluon
contribution. The relative difference between these fake rates is assigned as a
systematic uncertainty, which contributes an uncertainty of about 8\% to the
total background at $m_{\ell\tau}=1$~\si{\TeV}.

\subsubsection{$e\tau$ and $\mu\tau$ channels: multijet estimate}
\label{sec:multijet_estimate}

The multijet background contributions in the $e\tau$ and $\mu\tau$ channels are
evaluated using events in three control regions ($R_1$, $R_2$, $R_3$). The
events must pass the selection for the signal region, except that in $R_1$ and
$R_3$ the electron/muon must fail isolation and the
$\tau_\mathrm{had \mhyphen vis}$ candidate must fail identification, and in $R_1$
and $R_2$ the leptons must have the same electric charge and the electron/muon
$p_{\mathrm{T}}$ must be less than 200~\GeV to avoid signal  contamination.
For a \Zprime~boson with a mass of 500~\GeV, the lowest signal mass
considered in this paper, the contamination from the signal processs in $R_2$
is found to be below 1\%. The region definitions are summarized in
Table~\ref{tab:QCDRegions}. The background contribution is estimated as
$N_{\mathrm{MJ}} = N_{R_3} \times N_{R_2} / N_{R_1}$. The transfer factor
$N_{R_2}/N_{R_1}$ is calculated as a function of the dilepton mass to encapsulate
correlations between $m_{\ell\tau}$ and the isolation and identification
requirements, and it is fitted with a polynomial function. In each of the
regions defined, the contributions from other SM processes, such as~\W +jets,
\Z +jets, $\Z /\gamma^* \to \ell\ell$, diboson, and top-quark production, are
subtracted using simulation. The contribution from the multijet background is
$\sim$60\% ($\sim$20\%) of the \W+jets background for the $e\tau$ ($\mu\tau$)
channel, corresponding to $\sim$25\% ($\sim$10\%) of the total expected
background.

\begin{table}[!htbp]
  \centering
  \small
  \caption{Definition of the regions used for the multijet background
    estimation in the $e\tau$ and $\mu\tau$ channels.}
  \begin{tabular}{c|c|c}
    \hline
    & Object selection  & Lepton-pair charges \\
    \hline
    $R_1$ & Non-isolated $e/\mu$ \& 
    $\tau_\mathrm{had \mhyphen vis}$ failing $\tau$ ID requirements ($p_{T_{\ell}}$ \& $p_{T_{\tau}} <$ 200 GeV) & Same-charge\\
    $R_2$ & Isolated $e/\mu$ \&
    pass ID $\tau_\mathrm{had \mhyphen vis}$ ($p_{T_{\ell}}$ \& $p_{T_{\tau}} <$ 200 GeV) & Same-charge \\
    $R_3$ & Non-isolated $e/\mu$ \& $\tau_\mathrm{had \mhyphen vis}$
    failing $\tau$ ID requirements & Same-charge + Opposite-charge \\
    \hline
  \end{tabular}
  \label{tab:QCDRegions}
\end{table}

The multijet background is estimated using a transfer factor obtained using
same-charge lepton pairs and applied to opposite-charge plus same-charge lepton
pairs. To check the validity of this procedure, the multijet background is also
estimated using a transfer factor obtained with opposite-charge pairs. The
difference between the resulting shape and transfer factors is assigned as a
systematic uncertainty. The impact of this uncertainty is about 7\% at
\SI{1}{\TeV}.

The statistical uncertainties in the $m_{\ell\tau}$-dependent transfer factor and
the subtraction of simulated events are propagated to the final estimate and
assigned as a systematic uncertainty. The overall uncertainty is 50\% (15\%) at
\SI{1}{\TeV} for the $e\tau$ ($\mu\tau$) channel. The uncertainty in the
$\mu\tau$ channel is smaller because the transfer factor is found to have a
negligible effect on the dilepton invariant mass, and the transfer-factor fit
uncertainties are reduced.

\subsection{Reducible background validation}
\label{sec:bkg_summary}

\begin{figure}[!hbtp]
  \centering
  \subfloat[]{
    \includegraphics[width=0.49\columnwidth]{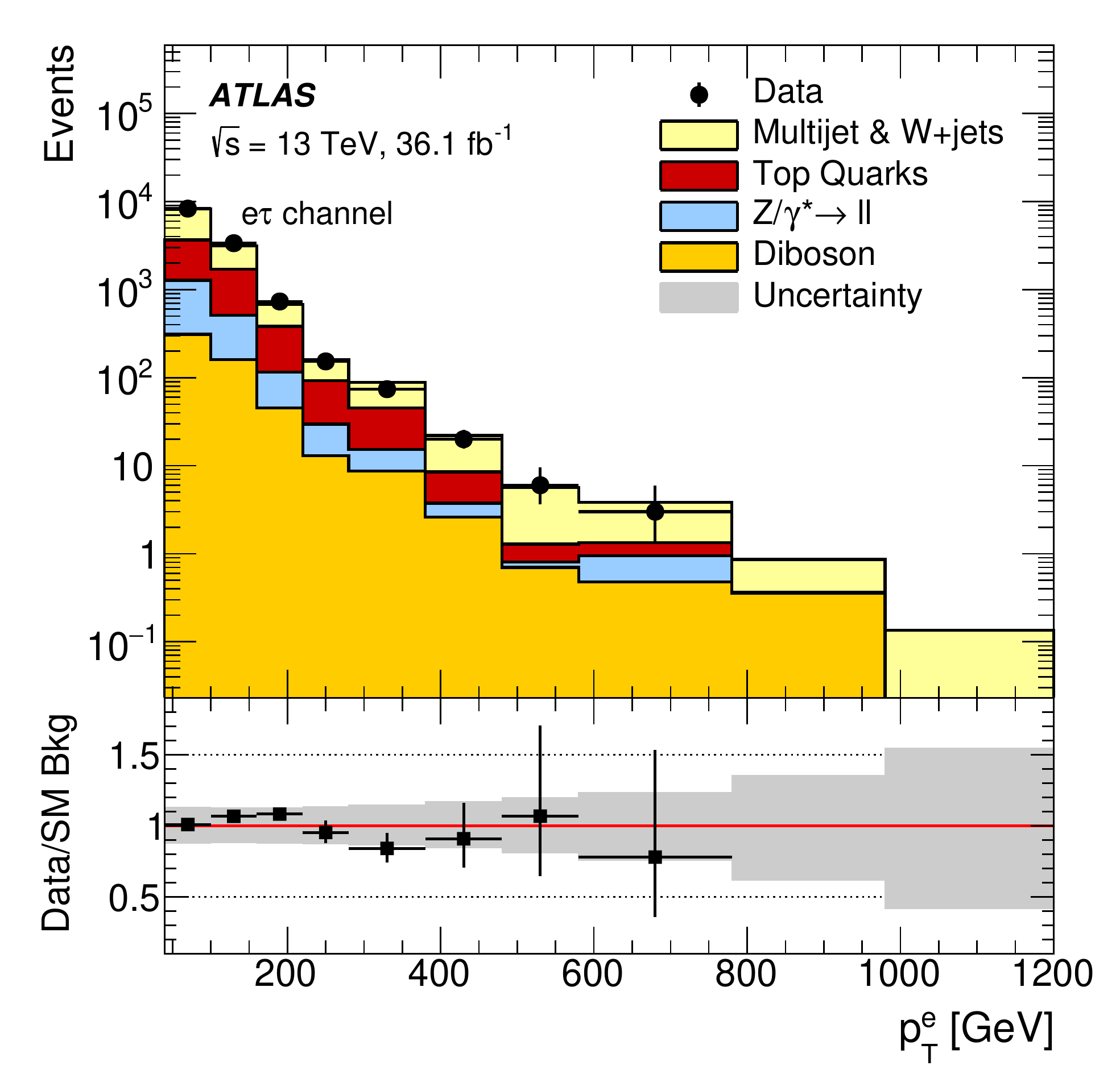}
  }
  \subfloat[]{
    \includegraphics[width=0.49\columnwidth]{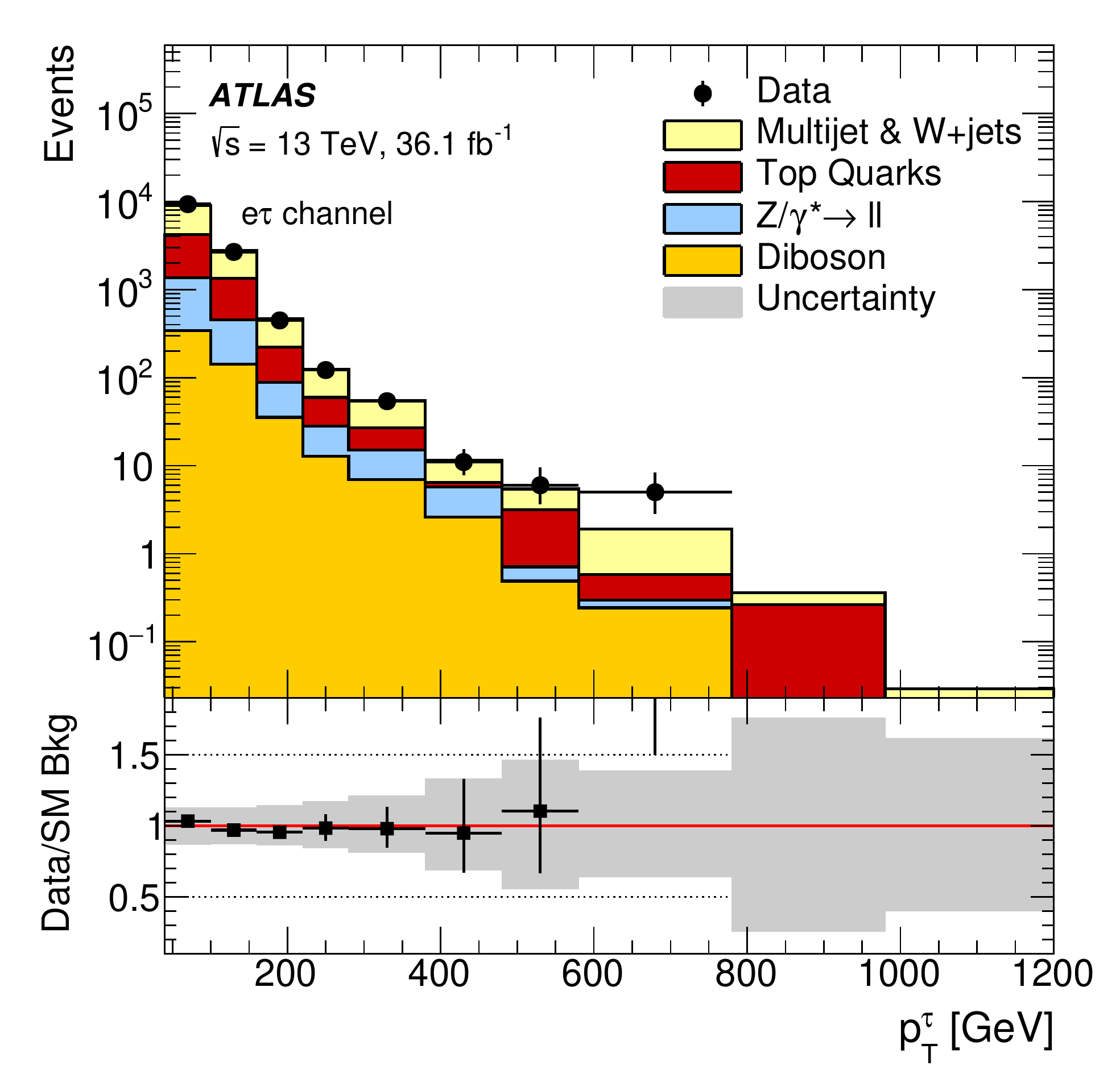}
  }\\
  \subfloat[]{
    \includegraphics[width=0.49\columnwidth]{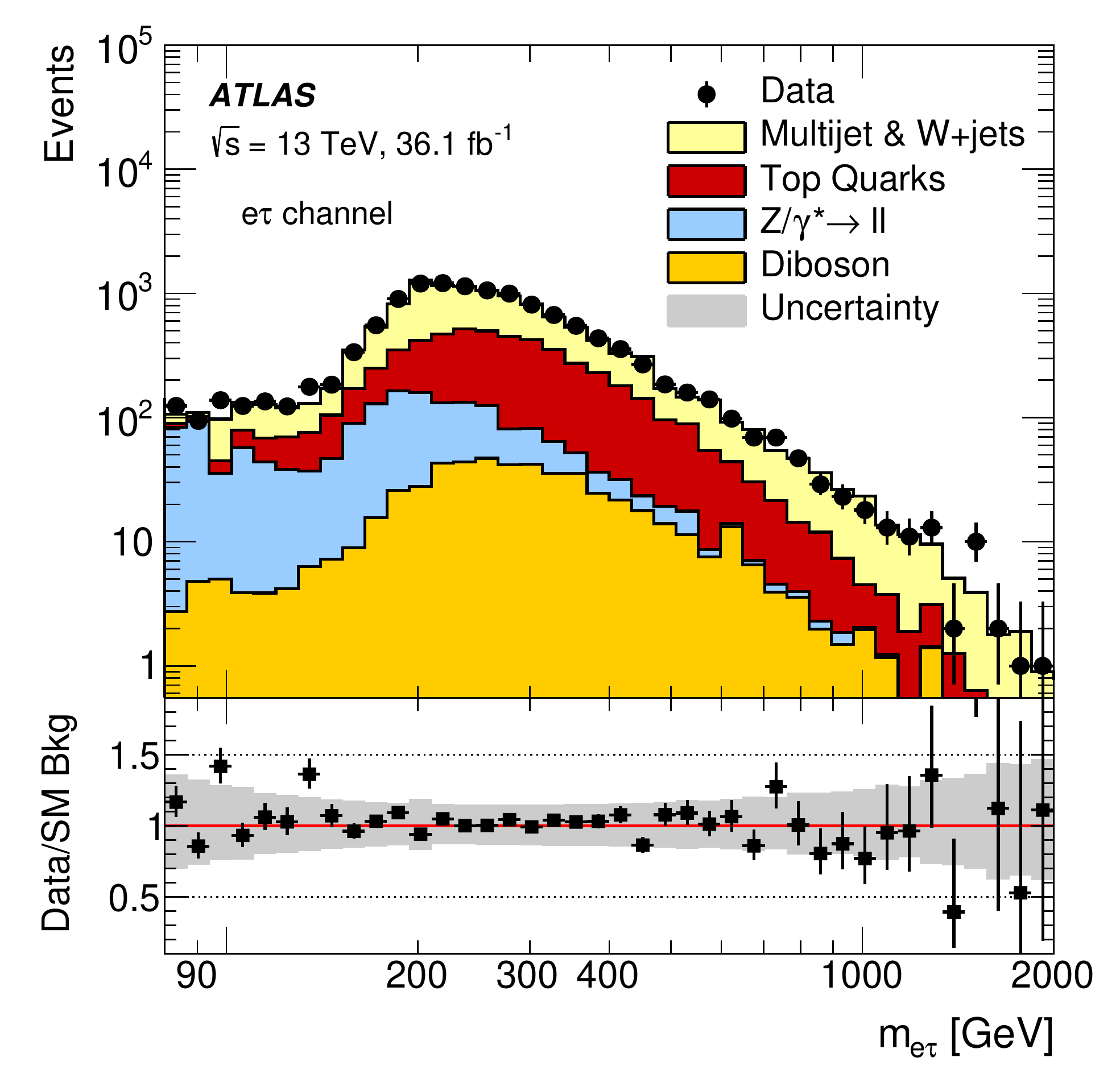}
  }
  \subfloat[]{
    \includegraphics[width=0.49\columnwidth]{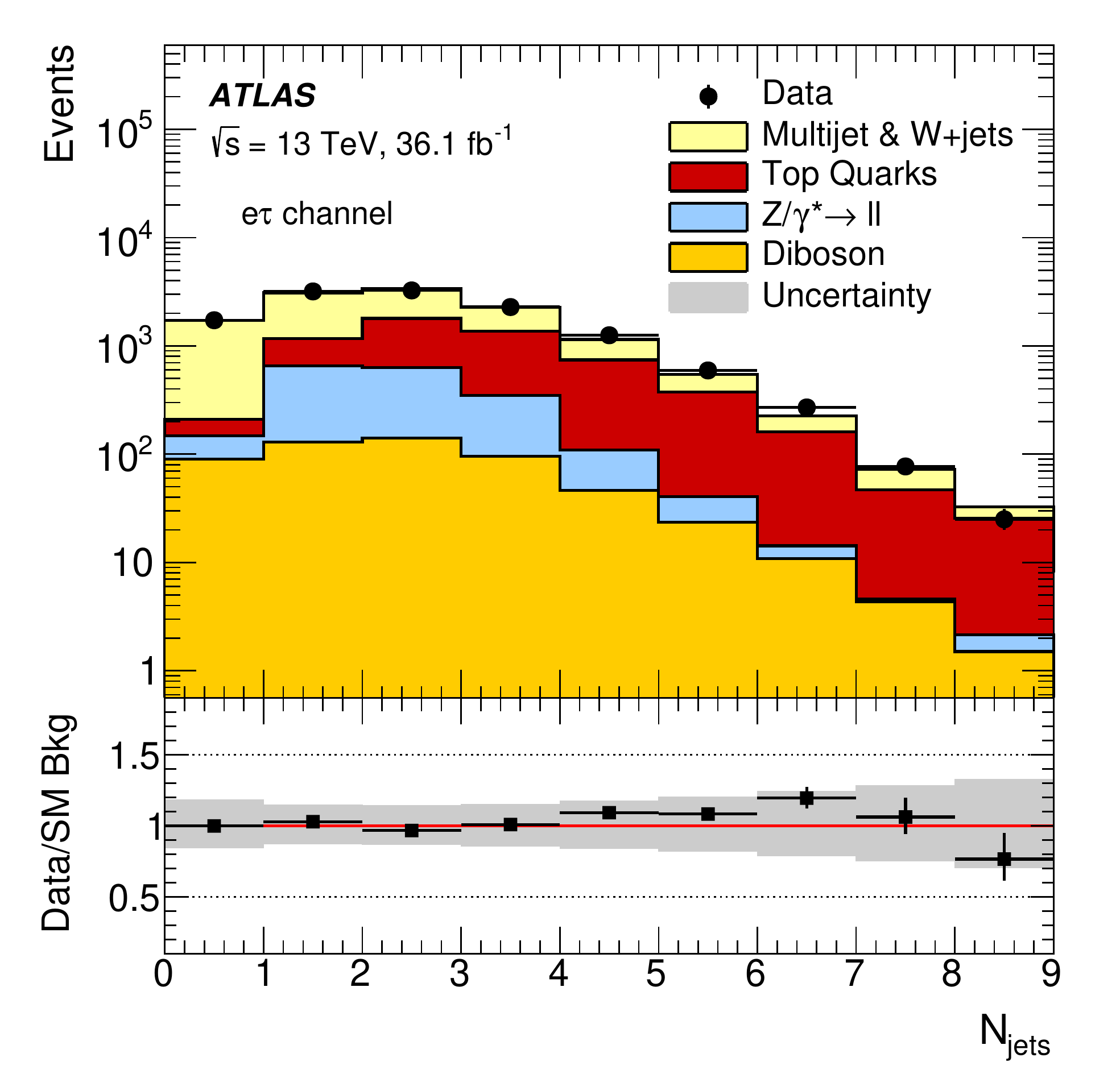}
  }
  \caption{Distributions in the \W+jets-enriched control region for the $e\tau$
    channel: (a) the electron and (b) $\tau$-lepton transverse momentum, (c)
    the $e\tau$ invariant mass, and (d) the jet multiplicity. No further data
    points are found in overflow bins.}
  \label{fig:background1}
\end{figure}

\begin{figure}[!hbtp]
  \centering
  \subfloat[]{
    \includegraphics[width=0.49\columnwidth]{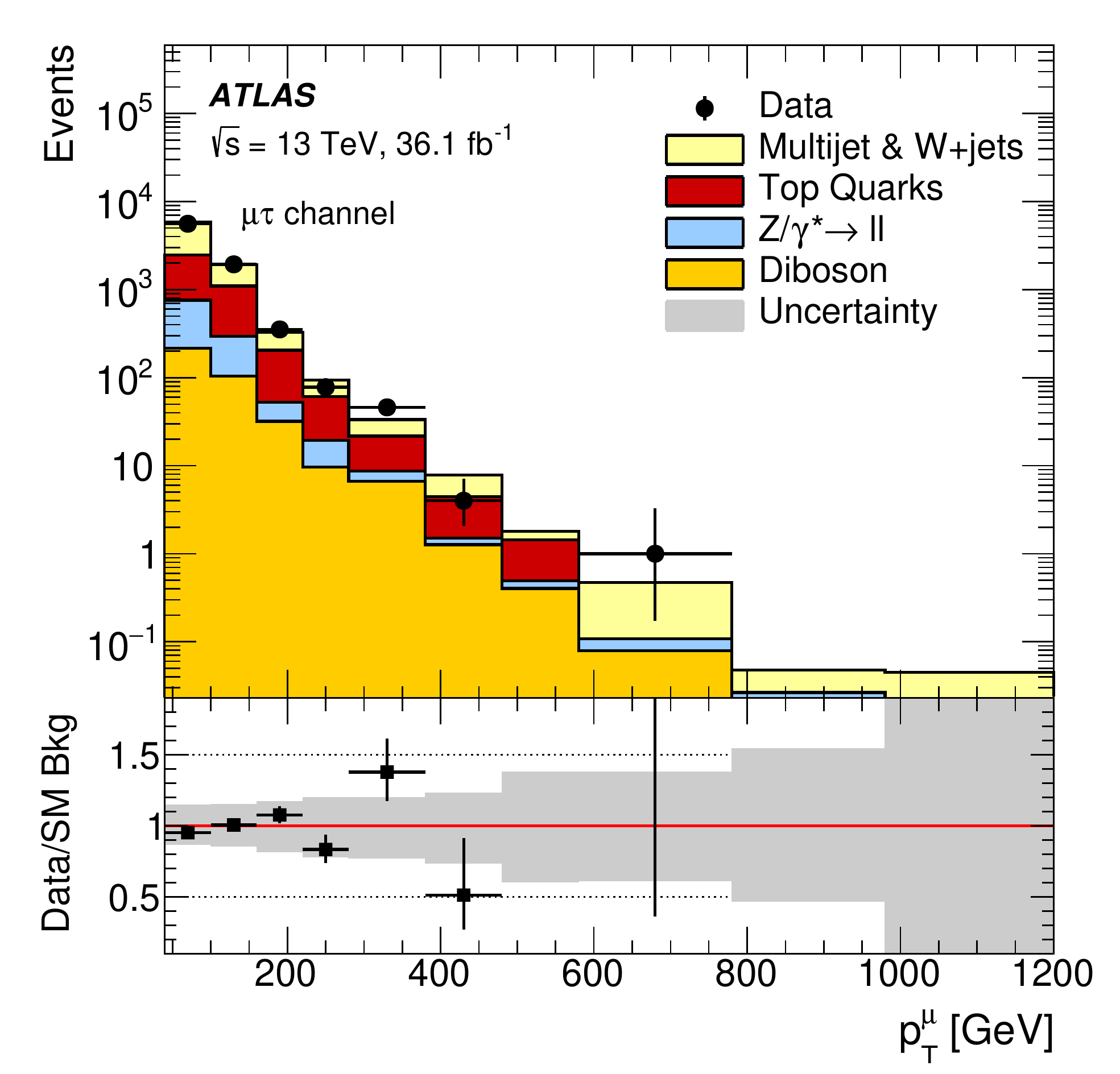}
  }
  \subfloat[]{
    \includegraphics[width=0.49\columnwidth]{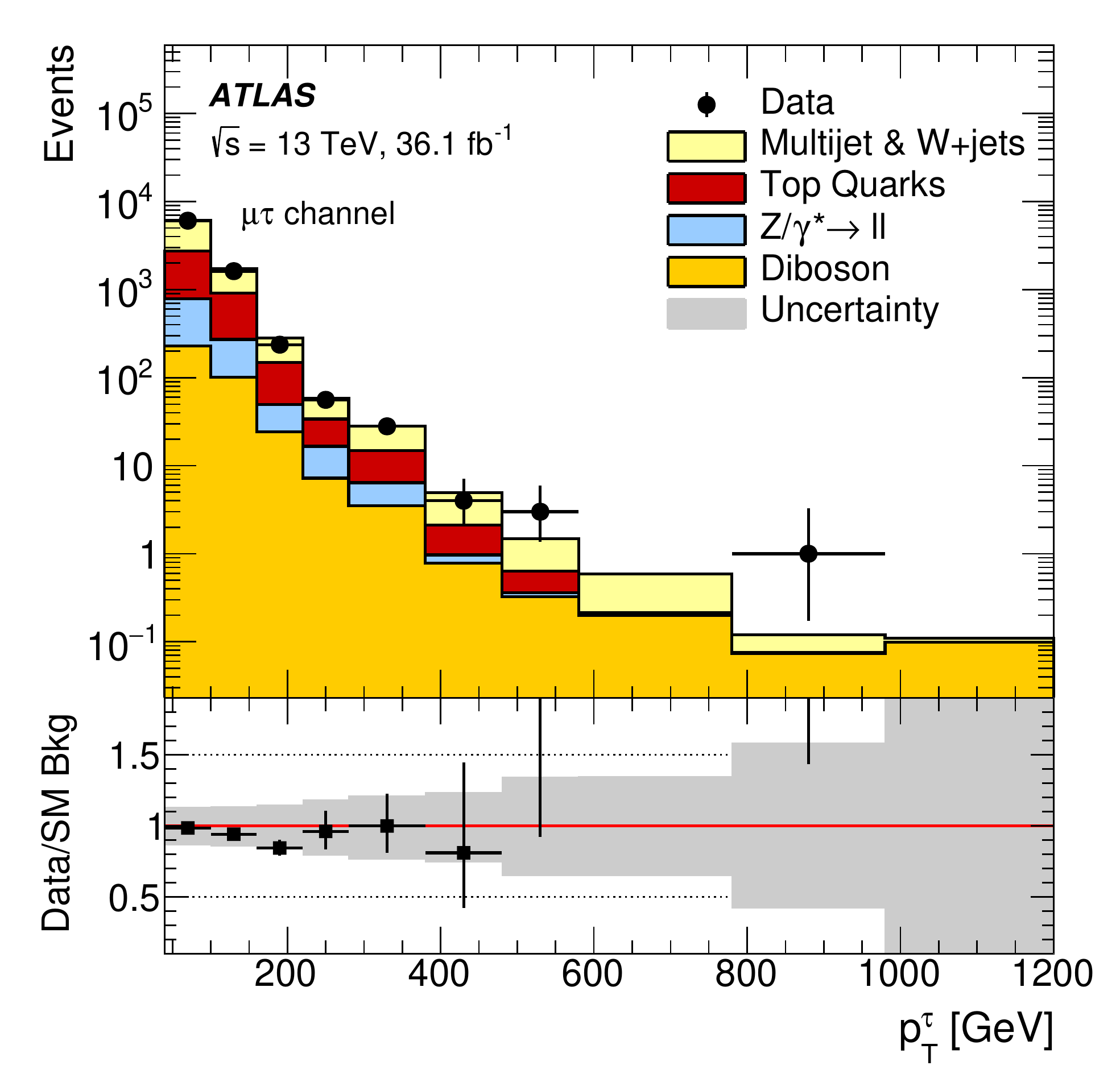}
  }\\
  \subfloat[]{
    \includegraphics[width=0.49\columnwidth]{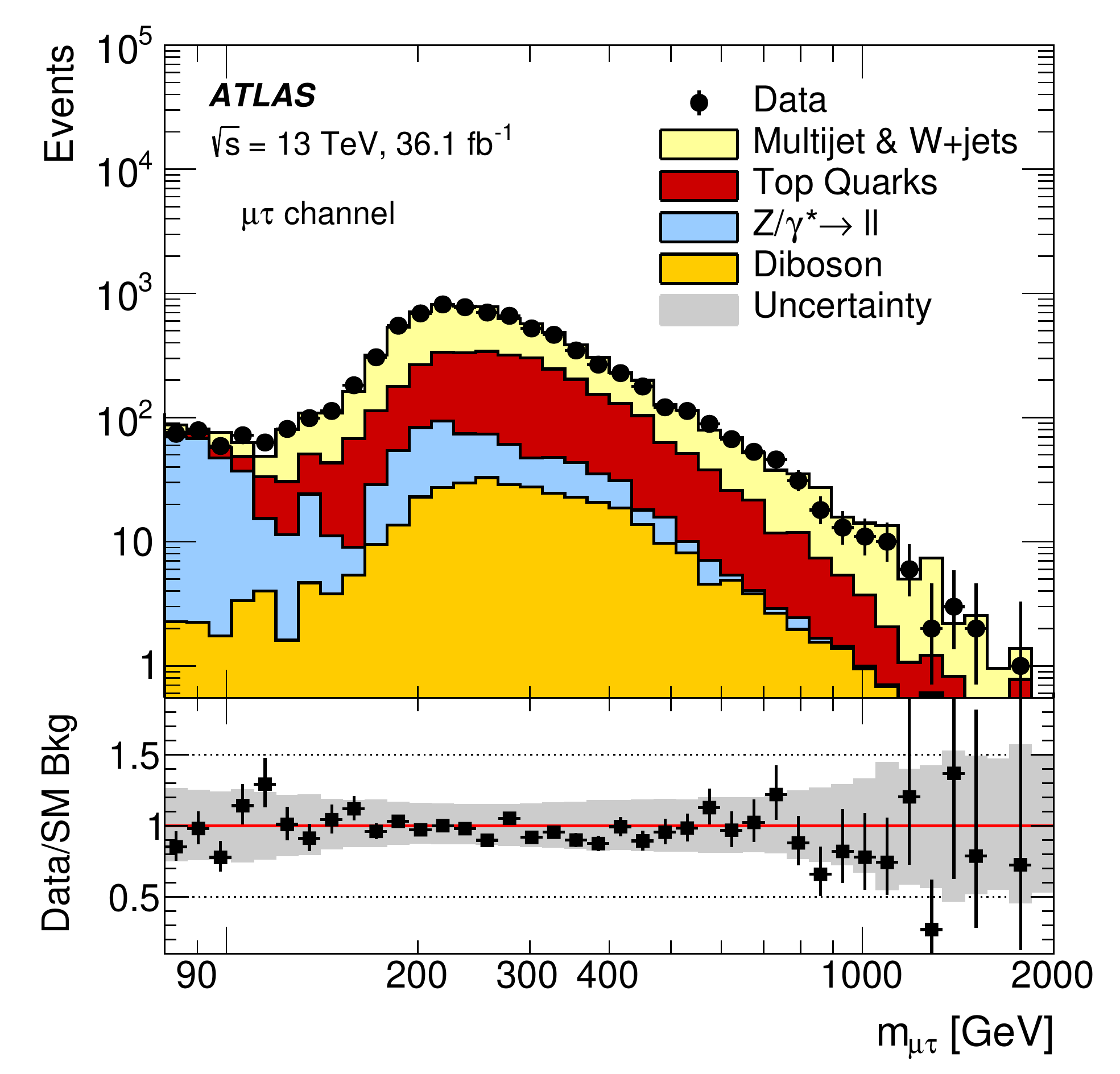}
  }
  \subfloat[]{
    \includegraphics[width=0.49\columnwidth]{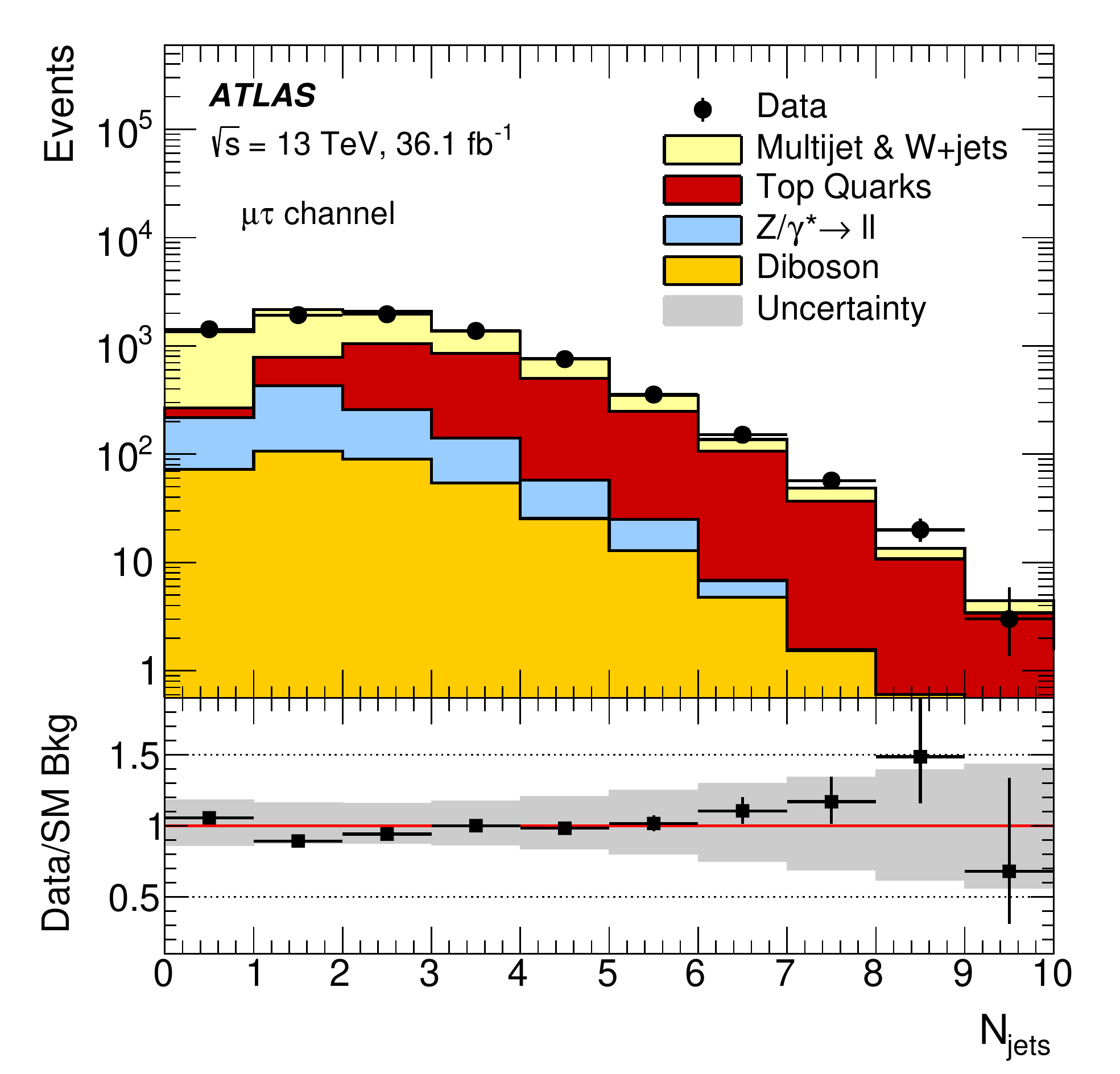}
  }
  \caption{Distributions in the \W+jets-enriched control region for the
     $\mu\tau$ channel: (a) the muon and (b) $\tau$-lepton transverse momentum,
     (c) the $\mu\tau$ invariant mass, and (d) the jet multiplicity. No further
     data points are found in overflow bins.}
  \label{fig:background2}
\end{figure}

The validity of the background estimation is checked in the \W+jets control region.
Figures~\ref{fig:background1} and \ref{fig:background2} show the electron, muon,
and $\tau$-lepton transverse momentum, $\ell\tau$ invariant mass and jet
multiplicity distributions for the $e\tau$ and $\mu\tau$ channels, respectively,
in the \W+jets control region. Good agreement is observed between the data and
the background prediction. The contribution from each SM background for each of
the final states in the signal region is given in Section~\ref{sec:results}.

\section{Systematic uncertainties}
\label{sec:systematics}
The sources of experimental uncertainty considered are pileup effects; lepton
efficiencies due to triggering, identification, reconstruction, isolation,
energy scale, and resolution
\cite{ATLAS-CONF-2016-024,PERF-2016-01,ATLAS-CONF-2017-029,PERF-2015-10,ATL-PHYS-PUB-2016-015};
jet energy scale and resolution~\cite{PERF-2016-04}; $b$-tagging
\cite{ATL-PHYS-PUB-2015-022}; and missing transverse momentum
\cite{bib:METperfo}. Sources of uncertainty  are considered for both
the simulated background and signal processes.

Mismodeling of the muon momentum resolution at the \si{\TeV} scale
from residual misalignment of the muon precision chambers can alter the
signal and background shapes. A corresponding uncertainty is obtained from
studies performed in dedicated data-taking periods with no magnetic field in
the MS. The muon reconstruction efficiency is affected at high $p_{\mathrm{T}}$ by
possible large energy losses in the calorimeter. The associated uncertainty is
estimated by comparing studies of $\Z\to\mu\mu$ data events extrapolated
to high $p_{\mathrm{T}}$ with the results predicted by simulation
\cite{EXOT-2015-07}. The effect on the muon reconstruction efficiency was
found to be approximately 3\% per \si{\TeV} as a function of muon
$p_{\mathrm{T}}$.

The uncertainty of the electron identification efficiency extrapolation is
determined from the differences in the electron shower shapes in the EM
calorimeters between data and simulation in the $\Z\to ee$ peak, which are
propagated to the high $p_{\mathrm{T}}$ electron sample. The effect on the
electron identification efficiency is 2\% and is independent of
$p_{\mathrm{T}}$ for electrons with transverse momentum above \SI{150}{\GeV}
\cite{EXOT-2015-07}.

The treatment of systematic uncertainties for $\tau$-leptons with
$p_{\mathrm{T}}$ up to \SI{100}{\GeV} is detailed in
Ref.~\cite{bib:TauReconstructionRun2}. An additional uncertainty of 20\% per
\si{\TeV} is assigned to the reconstruction efficiency of $\tau$-leptons
with $p_{\mathrm{T}}>100$~\si{\GeV} to account for the degradation of the modeling
and reconstruction efficiency from track merging, derived from studies in
simulation and in dijet data events at \SI{8}{\TeV} \cite{PERF-2013-06}.

The missing transverse momentum uncertainty is derived from the uncertainties
of the momenta of physics objects and uncertainties of the soft term determined
by comparisons with simulation.

A mis-modeling of the dilepton $p_T$ variable is found in the \ttbar simulation.
After reweighting to data in a \ttbar control region, an uncertainty is
assigned to account for the effect on the dilepton invariant mass spectrum.

A pile-up modeling uncertainty is estimated by varying the distribution of
pile-up events in the reweighting of the MC, to cover the uncertainty on the
ratio between the predicted and measured inelastic cross-section in the
fiducial volume defined by $M_{X}>13$~\si{\GeV} where $M_{X}$ is the mass of the
hadronic system \cite{STDM-2015-05}.

The uncertainty of 2.1\% in the luminosity applies to the signal and to
backgrounds derived from simulations. 

The uncertainties of the reducible background estimation in the $e\mu$ channel,
and the $\tau$-lepton fake rate, the multijet transfer factor calculation, and
the top-quark extrapolation are presented in Section~\ref{sec:background}.

The PDF uncertainties are the dominant systematic uncertainties affecting the
background estimates, together with the uncertainty on the extrapolation
to estimate the top-quark background contribution at high mass.
The contribution from PDF uncertainties is estimated using different PDF sets
and eigenvector variations within a particular PDF set for the top-quark,
diboson, and \W+jets backgrounds. The CT10 PDF uncertainty due to eigenvector
variations is evaluated through the use of LHAPDF \cite{bib:LHAPDF2014}
following the prescriptions of Ref.~\cite{bib:PDF4LHC}. The uncertainty related
to the choice of PDF is evaluated by comparing the results with those from the
central value of other PDF sets: MMHT2014 \cite{bib:Motylinski2014sya}, NNPDF3.0
\cite{bib:Ball2014uwa}, and CT14 \cite{bib:Dulat2015}. PDF-related
uncertainties in the signal shape are not considered. The uncertainties of the
$m_{\ell\ell^{\prime}}$ modeling in \ttbar events are obtained using separate
simulated samples generated with the renormalization scale and
\textit{h$_{\mathrm{damp}}$} parameter  varied by factors of 2 and 1/2, and are
referred to as``Top scale'' in Table~\ref{tab:sig_syst}. These uncertainties
for W+jets are not considered as they are found to be small, given that this
background is mainly composed of real lepton (e or mu) and fake tau pairs.
For the diboson background prediction, the PDF systematic is the leading
uncertainty.

Experimental systematic uncertainties common to signal and background processes
are assumed to be correlated. The systematic uncertainties of the estimated SM
background and signal yields are summarized in Tables~\ref{tab:sig_syst} and
\ref{tab:sig_syst_signal}.
For signal processes, only experimental systematic uncertainties are considered.
The simulated samples contribute a 3\% statistical uncertainty to the overall
signal acceptance times efficiency.
\begin{table}[!htbp]
  \centering
  \caption{Summary of the systematic uncertainties taken into
    account for background processes. Values are provided for $m_{\ell\ell^{\prime}}$
    values of 1, 2 and \SI{3}{\TeV}. Uncertainties are quoted as a percentage
    of the total background. The ``-'' sign indicates that the systematic
    uncertainty is not applicable.}
  \footnotesize
  \begin{tabular}{l|rrrr|rrrr|rrrr}
    \hline \hline
    Source & \multicolumn{4}{c|}{\SI{1}{\TeV}} & \multicolumn{4}{c|}{\SI{2}{\TeV}} &
    \multicolumn{4}{c}{\SI{3}{\TeV}}\\
    \cline{2-13}
    & $e\mu$ & $e\mu$ & $e\tau$ & $\mu\tau$ & $e\mu$ & $e\mu$ & $e\tau$ & $\mu\tau$
    & $e\mu$ & $e\mu$ & $e\tau$ & $\mu\tau$\\
    & & $b$-jet &  &  & & $b$-jet & &
    & & $b$-jet &  & \\
    & & veto &  &  & & veto & &
    & & veto &  & \\
    \hline                                
    Luminosity                            & 2 & 2    &  2   &  2          &  2 & 2   &  2 &  2        & 2 &   2   & 2   &  2   \\
    Top-quark extrapolation             & 5 &  3     &  2   &  2        & 32  &  8   &  3 &  4      & 63 &  12   & 3   & 14   \\
    Top scale		                     & 7 &  6     &  7   &  8        & 40 & 15   &  1 & 14      & 65 &  15   & 3   & 27   \\
    PDF                                   & 16 & 15     & 12   & 14       & 32 & 34   & 17 & 20      & 51 &  69   & 16  & 53   \\
    Pile-up               		  & 1 &  1   &  3   &  7      & 9  &  6   &  3 & 13      & 32 &  12   & 2   & 17   \\
    Dilepton $p_{\mathrm{T}}$ modeling              & 7 &  4     &  2   &  1      & 11 & 5   &  0 &  1    & 15 &  6   & 0 &  4   \\
    \hline
    Electron iden. and meas.	          & 4  &  4   &      5 & -       &  4  &   8  &   6 & -    &  5  & 11  &    8  & -      \\
    Muon iden. and meas.              &  3 &   4    & -    &   4      &  7  &   7  & - &  16     &  17 & 10  & -     &  18   \\
    $\tau$ iden. and meas.               & - & -      &  2   &  2        & - & - &  1 &  1        & - & -     &  1 &  2   \\
    $\tau$ reconstruction eff.   	         & - & -      &   2  &  2        & - & -  &   1  &   1      & - & -     &   1 &  3   \\
    $\tau$ fake rate                        & - & -      &  6   &  9        & - & -    &  5 & 12       & - & -       &   2 & 12   \\
    Multijet transf. factor              & - & -      & 31   &  2        & - & -    & 53 &  0     & - & -       &  64   &  0   \\
    Reducible $e\mu$ estimation        & 2 & -      & -   &  -        & 2 & -    & - &  -     & 2 & -       &  -   &  -   \\
    \hline
    Jet eff. and resol.                  &  1  &   4    &   9  &   8     &  2  &  12 &  17 &  42    &  5  &  17   &    22 &  48   \\
    $b$-tagging                          & - &  3   &  -   &  -          & - &  2    &  -  &  -     & - & 2       &  -    & -    \\
    $E_{\mathrm{T}}^{\mathrm{miss}}$ resol. and scale & - & -      &  3   &  4        & - & -    & 5   &  6      & - & -       &   8   &  8   \\
    \hline
    Total                                & 19 & 20    & 37    & 25       & 62 & 45  & 61  & 62      & 110 &  79   &  73   & 91   \\
    \hline \hline
  \end{tabular}
  \label{tab:sig_syst}
\end{table}
\begin{table}[!htbp]
  \centering
  \caption{Summary of the systematic uncertainties taken into account for
  signal processes. Values are provided for $m_{\ell\ell^{\prime}}$ values of 1, 2
  and \SI{3}{\TeV}.  The ``-'' sign indicates that the systematic uncertainty
  is not applicable.}
  \footnotesize
  \begin{tabular}{l|rrrr|rrrr|rrrr}
    \hline \hline
    Source & \multicolumn{4}{c|}{\SI{1}{\TeV}} & \multicolumn{4}{c|}{\SI{2}{\TeV}} &
    \multicolumn{4}{c}{\SI{3}{\TeV}}\\
    \cline{2-13}
    & $e\mu$ & $e\mu$ & $e\tau$ & $\mu\tau$ & $e\mu$ & $e\mu$ & $e\tau$ & $\mu\tau$
    & $e\mu$ & $e\mu$ & $e\tau$ & $\mu\tau$\\
    & & $b$-jet &  &  & & $b$-jet & &
    & & $b$-jet &  & \\
    & & veto &  &  & & veto & &
    & & veto &  & \\
    \hline                                
    Luminosity                            & 2 & 2    &  2   &  2          &  2   & 2   &  2 &  2      & 2   &   2 & 2     &  2   \\
    Pile-up               		  & 1 & 1   &  4   &  3           &  3   &  3   &  4 & 3      & 2   &   2 & 3     & 4   \\
    \hline
    Electron iden. and meas.	          & 6 &  6     &  8   & -         &  7  &   7 &  13 & -       &  9  &   9  &  14  & -      \\
    Muon iden. and meas.                  & 5 &   5    & -    &   5       &  6   &   6  & -  &  7     &  7  &   7  & -     &  8   \\
    $\tau$ iden. and meas.                & - & -      &  8   &  6        & -    & -    &  10 &  8    & -   &   -  &  11  &  9   \\
    $\tau$ reconstruction eff.   	  & - & -      &  2  &  2         & -    & -    &  2 &  2     & -   &   -  &   3   &  3   \\
    \hline
    Jet eff. and resol.                  &  1  &   1    &   2  &   2     &  1   &  1    &  3 &  4    &  1 &  1   &   3   &  3   \\
    $b$-tagging                          & - &  1   &  -   &  -            & -    &  1    &  -  &  -   & -  &  1   &  -    & -    \\
    $E_{\mathrm{T}}^{\mathrm{miss}}$ resol. and scale & - & 2   &  3 &  2  & -    & -     & 2   &  2   & -  &  -   &   3   &  2   \\
    \hline
    Total                                & 8 & 8    & 13    & 9            & 10 & 10  & 17  & 12      & 12 &  12   &  19  & 14   \\
    \hline \hline
  \end{tabular}
  \label{tab:sig_syst_signal}
\end{table}

\section{Results}
\label{sec:results}
Tables~\ref{tab:emu_yields}--\ref{tab:mutau_yields} show the expected and
observed numbers of events in the low and high mass regions for each channel.
The $e\mu$ background is dominated by \ttbar and diboson events,
while \W+jets events are dominant for the $e\tau$ and $\mu\tau$ final states.

\begin{table}[htbp]
  \centering
  \caption{Expected and observed numbers of $e\mu$ events in the low (a) and
    high (b) mass regions after applying all selection criteria. The
    statistical and systematic uncertainties are quoted.}
  \footnotesize
  \subfloat[$N_{\mathrm{events}}$ with $m_{e\mu}<600$ \si{\GeV}]{
    \begin{tabular}{l|c|c}
      \hline\hline 
      Process          & $m_{e\mu}<300$ \si{\GeV} & $300<m_{e\mu}<600$ \si{\GeV}\\
      \hline
      Top                   &  8460 $\pm$ 60 $\pm$ 860
                            &  2770 $\pm$ 30 $\pm$ 380 \\
      Diboson               &  1500 $\pm$ 20 $\pm$ 130
                            &   493 $\pm$  9 $\pm$  57 \\
      \W+jets               &   550 $\pm$ 30 $\pm$ 190
                            &   214 $\pm$ 14 $\pm$  75 \\
      \DY                   &    90 $\pm$  6 $\pm$  12
                            &    19.5 $\pm$  1.6 $\pm$   3.2 \\
      \hline
      Total background      & 10600 $\pm$ 70 $\pm$ 980
                            &  3490 $\pm$ 40 $\pm$ 440 \\

      Data                  & 10353
                            &  3417 \\
      \hline \hline               
    \end{tabular}
  }\\
  \subfloat[$N_{\mathrm{events}}$ with $m_{e\mu}>600$ \si{\GeV}]{
    \begin{tabular}{l|c|c}
      \hline\hline 
      Process & $600<m_{e\mu}<1200$ \si{\GeV} & $1200<m_{e\mu}<2000$ \si{\GeV} \\
      \hline
      Top                   &  140 $\pm$  6 $\pm$ 27
                            &    4.6 $\pm$  0.7 $\pm$  2.7 \\
      Diboson               &   47.5 $\pm$  1.2 $\pm$  8.0
                            &    2.96 $\pm$  0.31 $\pm$  0.79 \\
      \W+jets               &   24.1 $\pm$  3.9 $\pm$  8.4
                            &    0.1 $\pm$  2.3 $\pm$  0.0 \\
      \DY                   &    1.31 $\pm$  0.07 $\pm$  0.27
                            &    0.07 $\pm$  0.01 $\pm$  0.02 \\
      \hline
      Total background      &  213 $\pm$  7 $\pm$ 37
                            &    7.7 $\pm$  2.4 $\pm$  2.8 \\
      \hline
      Data                  &  196
                            &    1 \\
      \hline \hline
      \multicolumn{2}{c}{} & \\
      \hline \hline
      Process & $2000<m_{e\mu}<3000$ \si{\GeV} & $m_{e\mu}>3000$ \si{\GeV}\\
      \hline
      Top                   &   0.28 $\pm$ 0.09 $\pm$ 0.32
                            &  (0.16 $\pm$ 0.08 $\pm$ 0.28) $\cdot$ 10$^{-1}$ \\
      Diboson               &   0.25 $\pm$ 0.10 $\pm$ 0.11
                            &  (0.44 $\pm$ 0.01 $\pm$ 0.56) $\cdot$ 10$^{-2}$ \\
      \W+jets               &  $<0.01$
                            &  $<0.001$                                       \\
      \DY                   &  (0.48 $\pm$ 0.03 $\pm$ 0.23) $\cdot$ 10$^{-2}$
                            &  (0.16 $\pm$ 0.02 $\pm$ 0.31) $\cdot$ 10$^{-3}$ \\
      \hline
      Total background      &   0.54 $\pm$ 0.13 $\pm$ 0.41
                            &  (0.21 $\pm$ 0.08 $\pm$ 0.30) $\cdot$ 10$^{-1}$ \\
      \hline
      Data                  &   0
                            &   0                                             \\
      \hline \hline    
    \end{tabular}
  }
  \label{tab:emu_yields}
\end{table}
\begin{table}[htbp]
  \centering
  \caption{Expected and observed numbers of $e\mu$ events in the low (a) and
    high (b) mass regions  after applying all selection criteria including the
    $b$-jet veto. The statistical and systematic uncertainties are quoted.}
  \footnotesize
  \subfloat[$N_{\mathrm{events}}$ with $m_{e\mu}<600$ \si{\GeV}]{
    \begin{tabular}{l|c|c}
      \hline\hline 
      Process          & $m_{e\mu}<300$ \si{\GeV} & $300<m_{e\mu}<600$ \si{\GeV}\\
      \hline
      Top                  &  1660 $\pm$ 20 $\pm$ 260
                           &   570 $\pm$ 10 $\pm$ 100 \\
      Diboson              &  1470 $\pm$ 20 $\pm$ 130
                           &   479 $\pm$  8 $\pm$  55 \\
      \W+jets              &   231 $\pm$ 18 $\pm$  87
                           &    87 $\pm$  8 $\pm$  33 \\
      \DY                  &    86 $\pm$  6 $\pm$  12
                           &    18.4 $\pm$  1.3 $\pm$   3.0 \\
      \hline
      Total background     &  3450 $\pm$ 30 $\pm$ 350
                           &  1150 $\pm$ 20 $\pm$ 150 \\
      \hline
      Data                 & 3411
                           & 1082 \\
      \hline \hline               
    \end{tabular}
  }\\
  \subfloat[$N_{\mathrm{events}}$ with $m_{e\mu}>600$ \si{\GeV}]{
    \begin{tabular}{l|c|c}
      \hline\hline 
      Process & $600<m_{e\mu}<1200$ \si{\GeV} & $1200<m_{e\mu}<2000$ \si{\GeV}\\
      \hline
      Top                   &  28.6 $\pm$ 1.7 $\pm$  8.9
                            &   0.72 $\pm$ 0.10 $\pm$  0.85  \\
      Diboson               &  45.9 $\pm$ 1.2 $\pm$  7.7 
                            &   2.85 $\pm$ 0.30 $\pm$  0.76  \\
     \W+jets  		    &  11.0 $\pm$ 2.8 $\pm$  4.3
                            &   0.1 $\pm$ 1.9 $\pm$  0.0  \\
      \DY                   &   1.27 $\pm$ 0.06 $\pm$  0.25
                            &  (0.70 $\pm$ 0.05 $\pm$  0.20) $\cdot$ 10$^{-1}$\\
      \hline
      Total background      &  87 $\pm$ 3 $\pm$ 15
                            &   3.7 $\pm$ 2.0 $\pm$  1.1 \\
      \hline
      Data                  &  83
                            &   0 \\
      \hline \hline
      \multicolumn{2}{c}{} & \\
      \hline \hline
      Process & $2000<m_{e\mu}<3000$ \si{\GeV} & $m_{e\mu}>3000$ \si{\GeV}\\
      \hline
      Top                   & (2.8 $\pm$ 0.8 $\pm$  5.5) $\cdot$ 10$^{-2}$
                            & (0.8 $\pm$ 0.4 $\pm$  2.3) $\cdot$ 10$^{-3}$ \\
      Diboson               &  0.25 $\pm$ 0.10 $\pm$  0.11
                            & (0.42 $\pm$ 0.01 $\pm$  0.51) $\cdot$ 10$^{-2}$ \\
      \W+jets  & $<0.001$
                            & $<0.001$                                       \\
      \DY                   & (0.46 $\pm$ 0.03 $\pm$  0.23) $\cdot$ 10$^{-2}$
                            & (0.14 $\pm$ 0.02 $\pm$  0.30) $\cdot$ 10$^{-3}$ \\
      \hline
      Total background      &  0.28 $\pm$ 0.10 $\pm$  0.14
                            & (0.52 $\pm$ 0.04 $\pm$  0.60) $\cdot$ 10$^{-2}$ \\
      \hline
      Data                  &  0
                            &  0 \\
      \hline \hline    
    \end{tabular}
  }
  \label{tab:emu_yields_bjet}
\end{table}
\begin{table}[htbp]
  \centering
  \caption{Expected and observed numbers of $e\tau$ events in the low (a) and
    high (b) mass regions  after applying all selection criteria. The
    statistical and systematic uncertainties are quoted.}
  \footnotesize
  \subfloat[$N_{\mathrm{events}}$ with $m_{e\tau}<600$ \si{\GeV}]{
    \begin{tabular}{l|c|c}
      \hline\hline 
      Process        & $m_{e\tau}<300$ \si{\GeV} & $300<m_{e\tau}<600$ \si{\GeV}\\
      \hline
      Top                   &  2020 $\pm$  30 $\pm$  390
                            &  1800 $\pm$  30 $\pm$  370 \\
      Diboson               &   465 $\pm$  10 $\pm$   77
                            &   330 $\pm$   8 $\pm$   58 \\
      Multijet and \W+jets  & 13200 $\pm$ 200 $\pm$ 2900
                            &  3100 $\pm$  70 $\pm$  870 \\
      \DY                   &  3300 $\pm$  60 $\pm$  500
                            &   610 $\pm$  20 $\pm$  130 \\
      \hline
      Total background      & 19000 $\pm$ 200 $\pm$ 3300
                            &  5800 $\pm$  100 $\pm$ 1100 \\
      \hline
      Data                  & 19532
                            &  5858 \\
      \hline \hline               
    \end{tabular}
  }\\
  \subfloat[$N_{\mathrm{events}}$ with $m_{e\tau}>600$ \si{\GeV}]{
    \begin{tabular}{l|c|c}
      \hline\hline 
      Process & $600<m_{e\tau}<1200$ \si{\GeV} & $1200<m_{e\tau}<2000$ \si{\GeV}\\
      \hline
      Top                   & 161 $\pm$  2 $\pm$  53
                            &   4.6 $\pm$  0.4 $\pm$   2.2 \\
      Diboson               &  48 $\pm$  2 $\pm$  11
                            &   4.7 $\pm$  0.8 $\pm$   1.7 \\
      Multijet and \W+jets  & 300 $\pm$ 20 $\pm$ 140
                            &  24 $\pm$  2 $\pm$  16 \\
      \DY                   &  25.6 $\pm$  0.6 $\pm$   6.0
                            &   1.30 $\pm$  0.04 $\pm$   0.42 \\
      \hline
      Total background      & 540 $\pm$ 20 $\pm$ 160
                            &  34 $\pm$  2 $\pm$  17 \\
      \hline
      Data                  & 480
                            &  24 \\
      \hline \hline
      \multicolumn{2}{c}{} & \\
      \hline \hline
      Process & $2000<m_{e\tau}<3000$ \si{\GeV} & $m_{e\tau}>3000$ \si{\GeV}\\
      \hline
      Top                   &  0.13 $\pm$ 0.04 $\pm$ 0.10
                            & (0.20 $\pm$ 0.10 $\pm$ 0.32) $\cdot$ 10$^{-2}$ \\
      Diboson               &  0.41 $\pm$ 0.16 $\pm$ 0.21
                            & (0.35 $\pm$ 0.04 $\pm$ 0.39) $\cdot$ 10$^{-2}$ \\
      Multijet and \W+jets  &  2.4 $\pm$ 0.3 $\pm$ 2.0
                            &  0.30 $\pm$ 0.14 $\pm$ 0.19 \\
      \DY                   &  0.09 $\pm$ 0.02 $\pm$ 0.04
                            & (0.34 $\pm$ 0.01 $\pm$ 0.39) $\cdot$ 10$^{-2}$ \\
      \hline
      Total background      &  3.1 $\pm$ 0.4 $\pm$ 2.1
                            &  0.31 $\pm$ 0.15 $\pm$ 0.23 \\
      \hline
      Data                  &  5
                            &  0 \\
      \hline \hline    
    \end{tabular}
  }
  \label{tab:etau_yields}
\end{table}

\begin{table}[htbp]
  \centering
  \caption{Expected and observed numbers of $\mu\tau$ events in the low (a) and
    high (b) mass regions after applying all selection criteria. The
    statistical and systematic uncertainties are quoted.}
  \footnotesize
  \subfloat[$N_{\mathrm{events}}$ with $m_{\mu\tau}<600$ \si{\GeV}]{
    \begin{tabular}{l|c|c}
      \hline\hline 
      Process          & $m_{\mu\tau}<300$ \si{\GeV} & $300<m_{\mu\tau}<600$ \si{\GeV}\\
      \hline
      Top                   &  1380 $\pm$  20 $\pm$  300
                            &  1160 $\pm$  20 $\pm$  250 \\
      Diboson               &   318 $\pm$   8 $\pm$   55
                            &   225 $\pm$   6 $\pm$   42 \\
      Multijet and \W+jets  &  6900 $\pm$ 200 $\pm$ 1400
                            &  1650 $\pm$  50 $\pm$  380 \\
      \DY                   &  1650 $\pm$  40 $\pm$  270
                            &   339 $\pm$  14 $\pm$   71 \\
      \hline
      Total background      & 10300 $\pm$ 200 $\pm$ 1700
                            &  3380 $\pm$  60 $\pm$  550 \\
      \hline
      Data                  & 10525
                            &  3378 \\
      \hline \hline               
    \end{tabular}
  }\\
  \subfloat[$N_{\mathrm{events}}$ with $m_{\mu\tau}>600$ \si{\GeV}]{
    \begin{tabular}{l|c|c}
      \hline\hline 
      Process & $600<m_{\mu\tau}<1200$ \si{\GeV} & $1200<m_{\mu\tau}<2000$ \si{\GeV}\\
      \hline
      Top                   &  95 $\pm$  1 $\pm$ 26
                            &   3.5 $\pm$  0.2 $\pm$  2.7 \\
      Diboson               &  33.3 $\pm$  1.7 $\pm$  9.2
                            &   2.9 $\pm$  0.5 $\pm$  1.5 \\
      Multijet and \W+jets  & 140 $\pm$  10 $\pm$ 43
                            &   6.4 $\pm$  1.0 $\pm$  2.4 \\
      \DY                   &  14.4 $\pm$  1.2 $\pm$  4.3
                            &   0.88 $\pm$  0.07 $\pm$  0.32 \\
      \hline
      Total background      & 282 $\pm$  10 $\pm$ 61
                            &  13.7 $\pm$  1.1 $\pm$  5.0 \\
      \hline
      Data                  & 255
                            &  12 \\
      \hline \hline
      \multicolumn{2}{c}{} & \\
      \hline \hline
      Process & $2000<m_{\mu\tau}<3000$ \si{\GeV} & $m_{\mu\tau}>3000$ \si{\GeV}\\
      \hline
      Top                   &  0.17 $\pm$ 0.03 $\pm$ 0.21
                            & (0.56 $\pm$ 0.19 $\pm$ 0.94) $\cdot$ 10$^{-2}$ \\
      Diboson               &  0.54 $\pm$ 0.30 $\pm$ 0.38
                            & (0.62 $\pm$ 0.09 $\pm$ 0.95) $\cdot$ 10$^{-2}$ \\
      Multijet and \W+jets  &  0.87 $\pm$ 0.35 $\pm$ 0.89
                            & (0.87 $\pm$ 0.70 $\pm$ 0.60) $\cdot$ 10$^{-2}$ \\
      \DY                   & (0.78 $\pm$ 0.03 $\pm$ 0.42) $\cdot$ 10$^{-1}$
                            & (0.63 $\pm$ 0.04 $\pm$ 0.84) $\cdot$ 10$^{-2}$ \\
      \hline
      Total background      &  1.65 $\pm$ 0.46 $\pm$ 1.30 
                            & (0.27 $\pm$ 0.07 $\pm$ 0.33) $\cdot$ 10$^{-1}$ \\
      \hline
      Data                  &  2
                            &  0 \\
      \hline \hline    
    \end{tabular}
  }
  \label{tab:mutau_yields}
\end{table}

\begin{figure}[!hbtp]
  \centering
  \subfloat[]{
    \includegraphics[width=0.49\columnwidth]{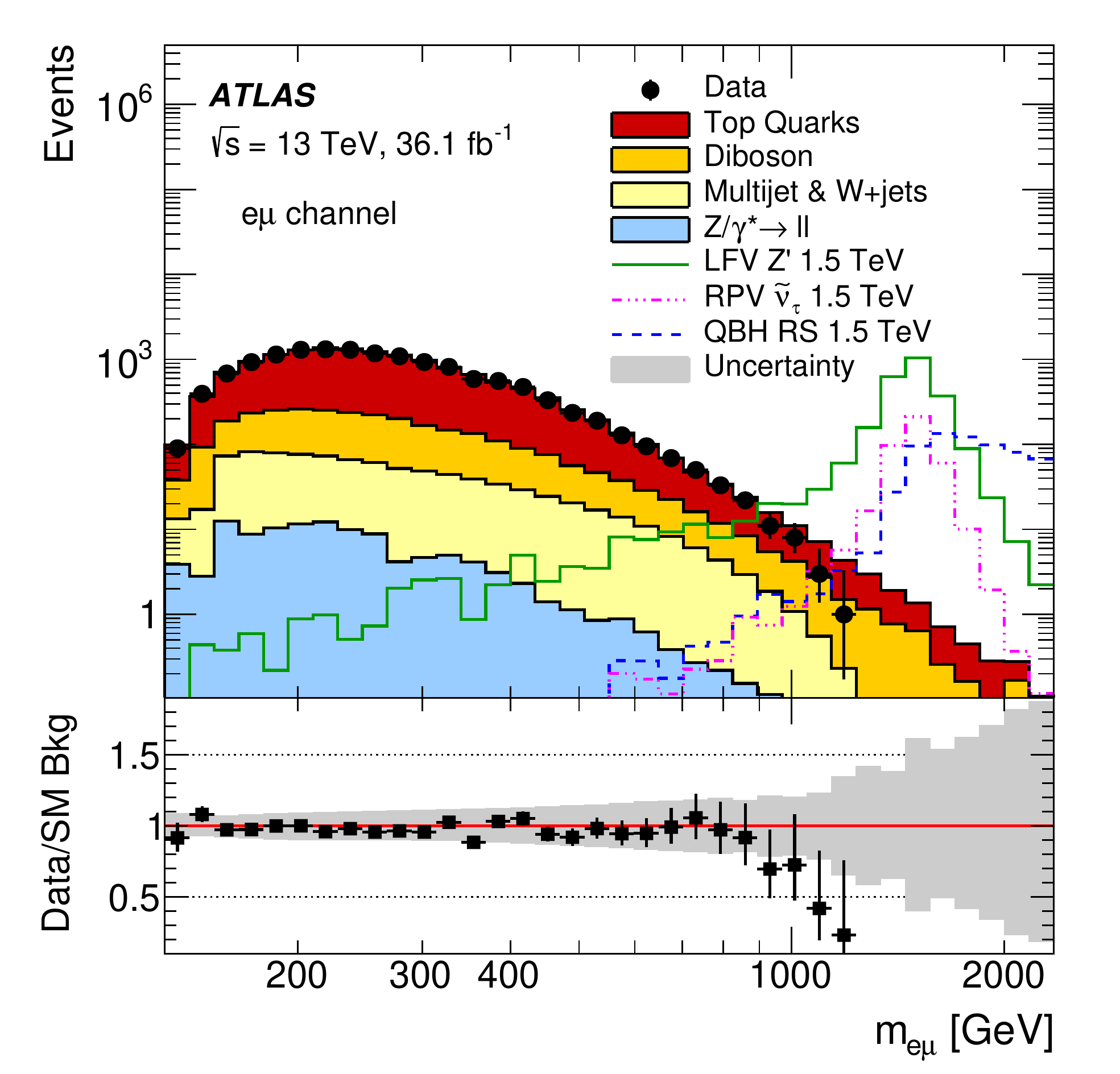}
  }
  \subfloat[]{
    \includegraphics[width=0.49\columnwidth]{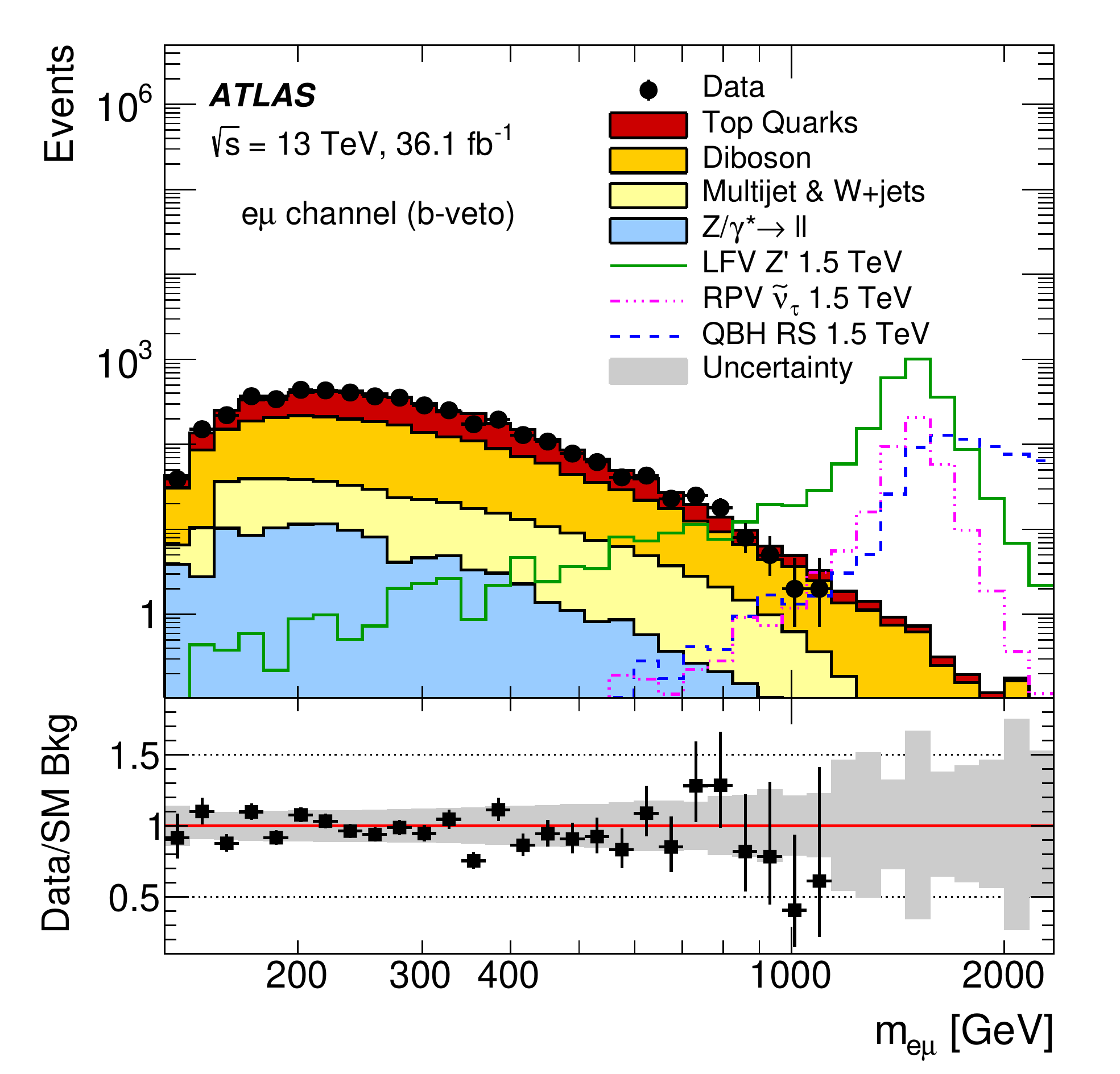}
  }\\
  \subfloat[]{
    \includegraphics[width=0.49\columnwidth]{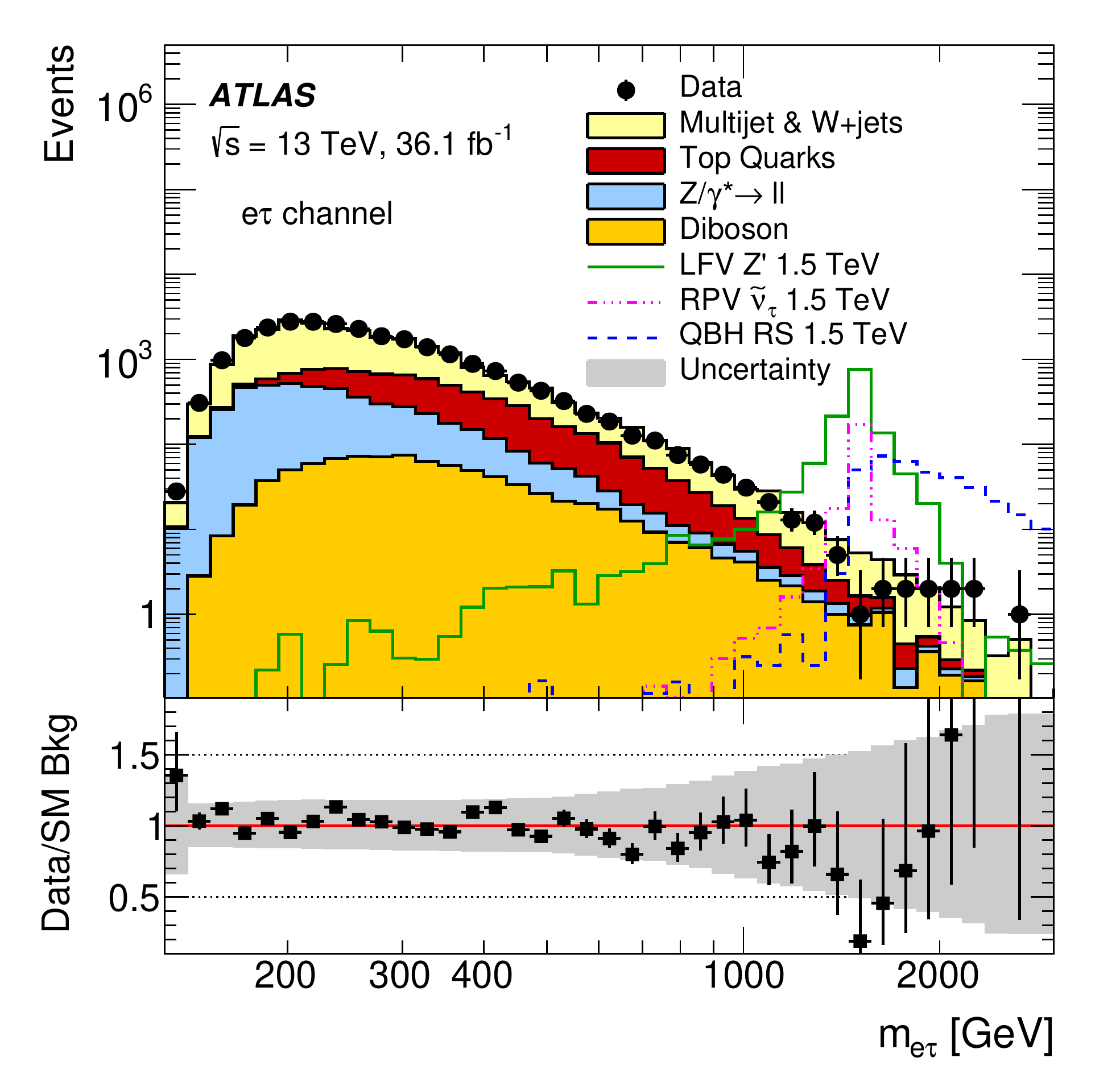}
  }
  \subfloat[]{
    \includegraphics[width=0.49\columnwidth]{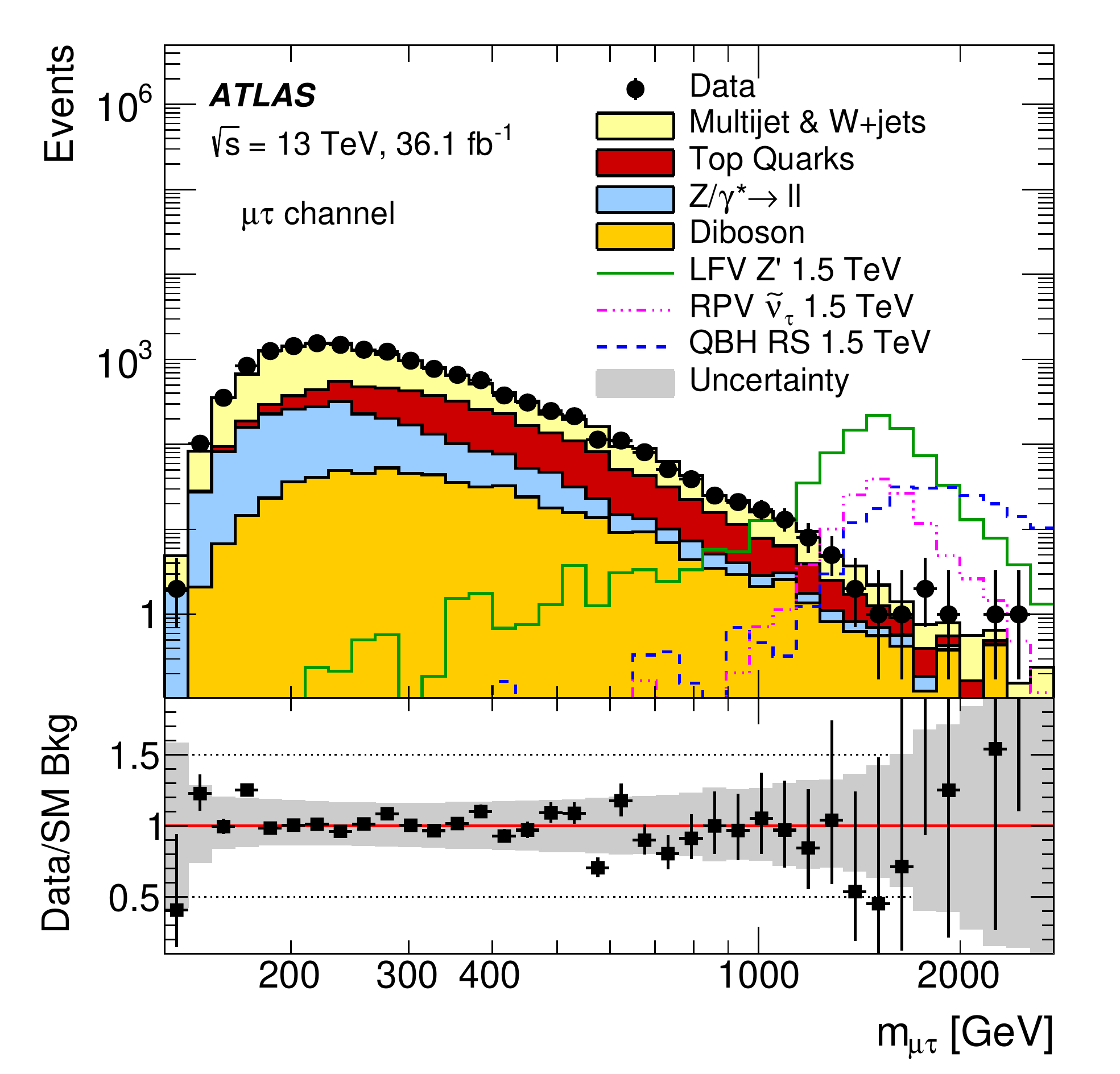}
  }
  \caption{The invariant mass distribution of (a) $e\mu$, (b) $e\mu$ with
    $b$-veto, (c) $e\tau$, and (d) $\mu\tau$ pairs for data and the SM
    predictions. Three signal examples are overlaid: a \Zprime~boson with a
    mass of \SI{1.5}{\TeV}, a $\tau$-sneutrino ($\tilde{\nu}_{\tau}$) with a
    mass of \SI{1.5}{\TeV}, and a RS quantum black-hole (QBH) with a threshold
    mass of \SI{1.5}{\TeV}. The range is chosen such that all data points are
    visible. The error bars show the Poissonian statistical uncertainty of the
    observed yields, while the band in the bottom plot includes all systematic
    uncertainties combined in quadrature. No further data points are found in
    overflow bins.}
  \label{fig:invmass}
\end{figure}

Figure~\ref{fig:invmass} shows the dilepton invariant mass distributions
for the $e\mu$, $e\mu$ with $b$-veto, $e\tau$, and $\mu\tau$ channels.
The largest deviation found in the data is a deficit in the 1.1--1.4~\si{\TeV}
range of the $e\mu$ channel, with a global significance of 1.8 standard
deviations, obtained using the BumpHunter program~\cite{Choudalakis:2011bh}.
Due to the parton luminosity tail in the LFV \Zprime~model, the impact of the
deficit in the 1.1--1.4~\si{\TeV} range is also seen in the observed limit for
\Zprime~boson masses up to 4--5~\si{\TeV}. No significant excess is found in
any channel.

The electron--muon event with an invariant mass of \SI{2.1}{\TeV} found in the
previous version of this analysis~\cite{EXOT-2015-20} no longer satisfies the
event selection, since the previously selected muon candidate is found to
overlap with a jet using the criteria of this paper and is no longer classified
as a prompt muon.

Since no deviations from the SM prediction are observed, model-dependent
exclusion limits are extracted using a Bayesian method implemented with
the Bayesian analysis toolkit \cite{bib:Caldwell2008}. A binned likelihood
function is constructed from the product of the Poisson probabilities of the
observed and expected numbers of events in each $m_{\ell\ell^\prime}$ mass bin
as in Ref.~\cite{EXOT-2015-20}. A 95\% credibility level (CL) Bayesian upper
limit is placed on the signal cross-section times branching ratio.

Expected exclusion limits are obtained by generating 1000 pseudo-experiments
for each signal mass point. The median value of the pseudo-experiment
distribution of the 95\% CL Bayesian upper limit is taken as the expected
limit. The one- and two-standard deviation intervals of the expected limit are
obtained by finding the 68\% and 95\% intervals of the pseudo-experiment upper
limit distribution, respectively.

The invariant mass spectrum for each final state is analyzed in 60 bins
from 120~\GeV\ to 10~\TeV. The bin width is around 7\% of the dilepton mass
throughout the whole range.
The predicted width of the \Zprime~boson, 3\% for $m_{\Zprime}=2$~\si{\TeV}, is
smaller than the detector resolutions for the $e\mu$ and the $\mu\tau$ channels,
which are approximately 8\% and 12\%, respectively, at the same \Zprime~boson
mass. For the $e\tau$ final state the detector resolution is 4\% at
$m_{\Zprime}=2$~\si{\TeV}, comparable to the \Zprime~boson width. The width of
the $\tilde{\nu}_{\tau}$ is below 1\%, and hence the resolution of the detector
is larger than the width for each of the final states investigated.

\begin{figure}[!htbp]
  \centering
  \subfloat[]{
    \includegraphics[width=0.49\linewidth]{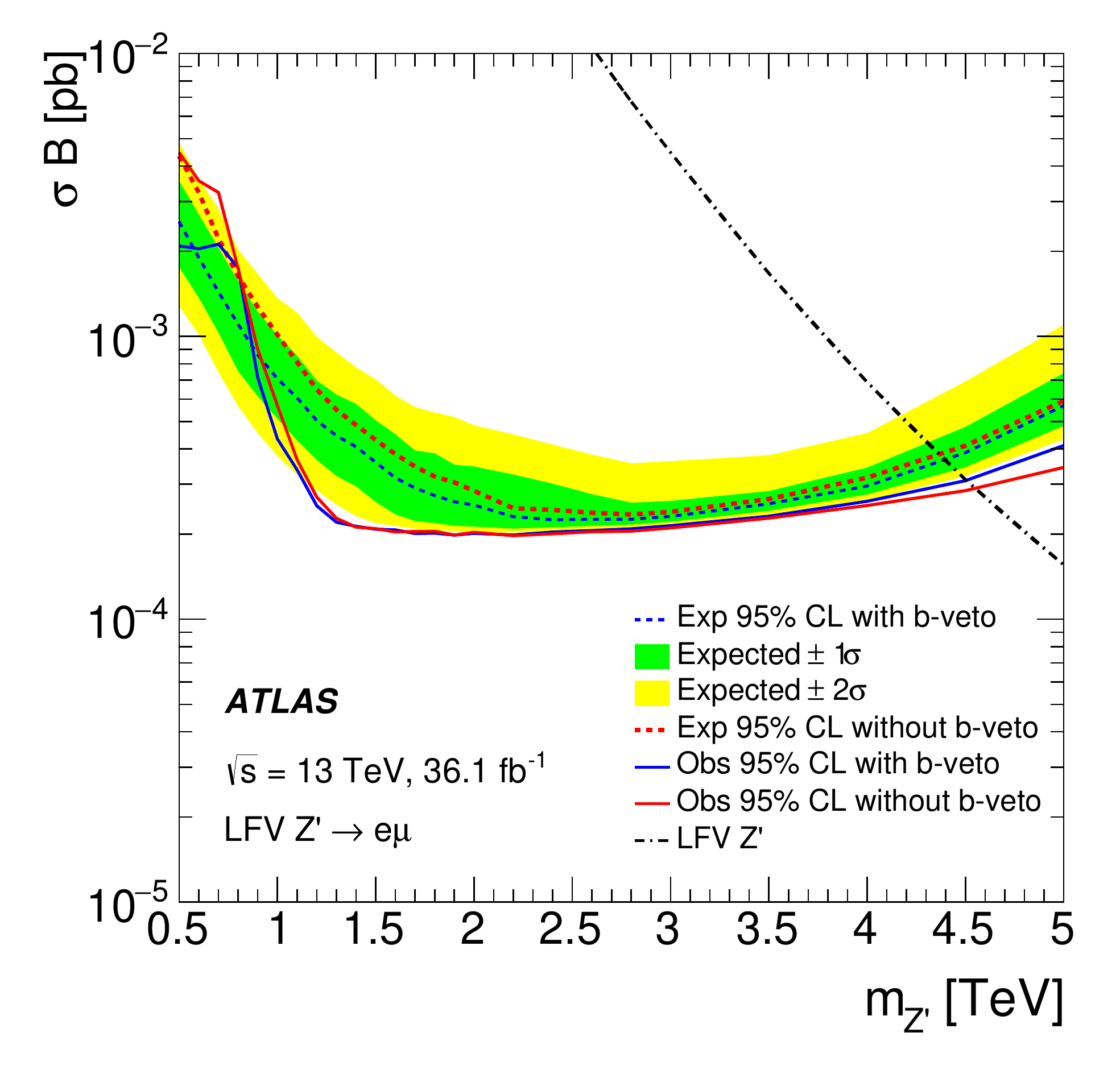}
  }
  \subfloat[]{
    \includegraphics[width=0.49\linewidth]{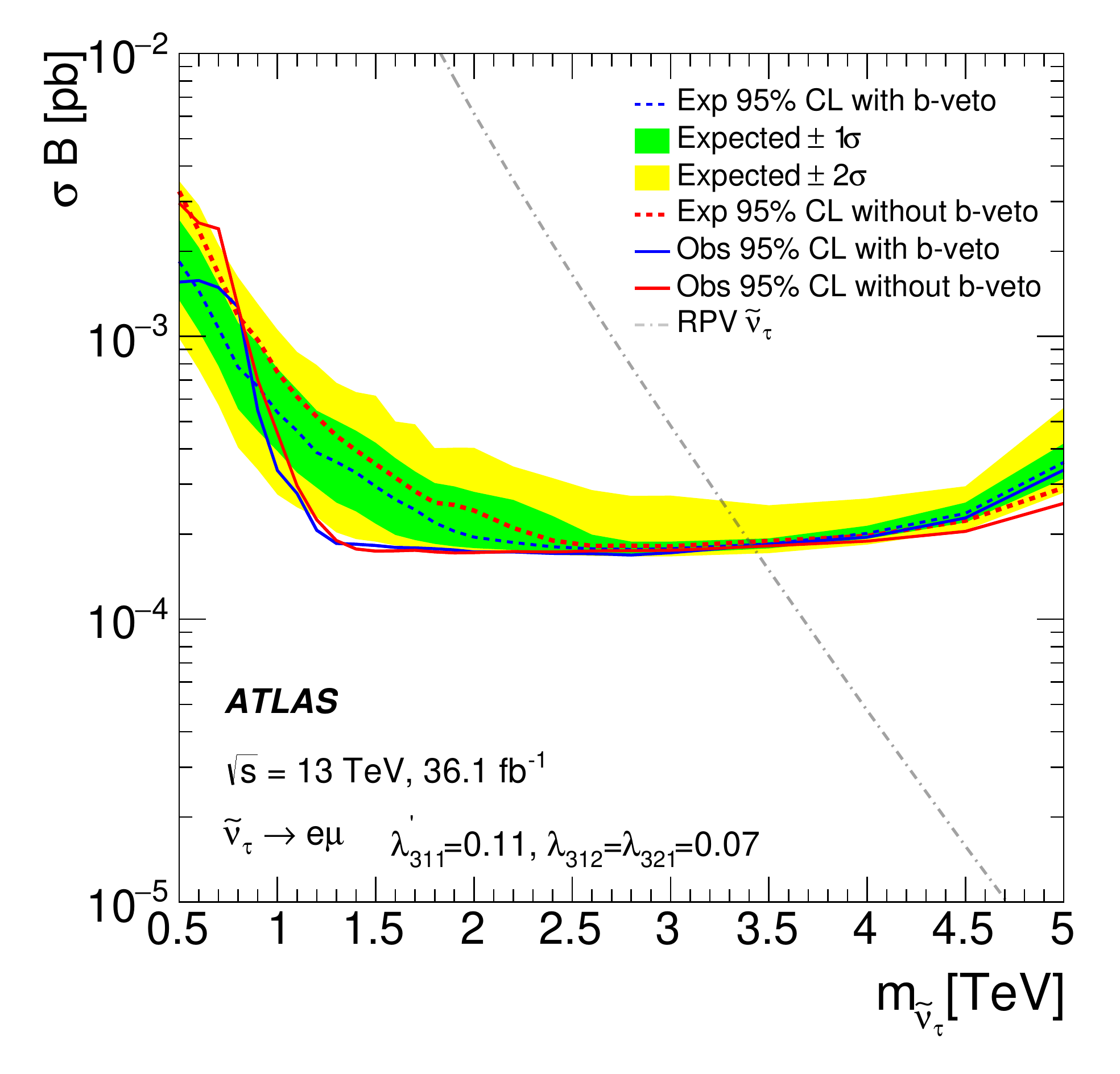}
  } \\
  \subfloat[]{
    \includegraphics[width=0.49\linewidth]{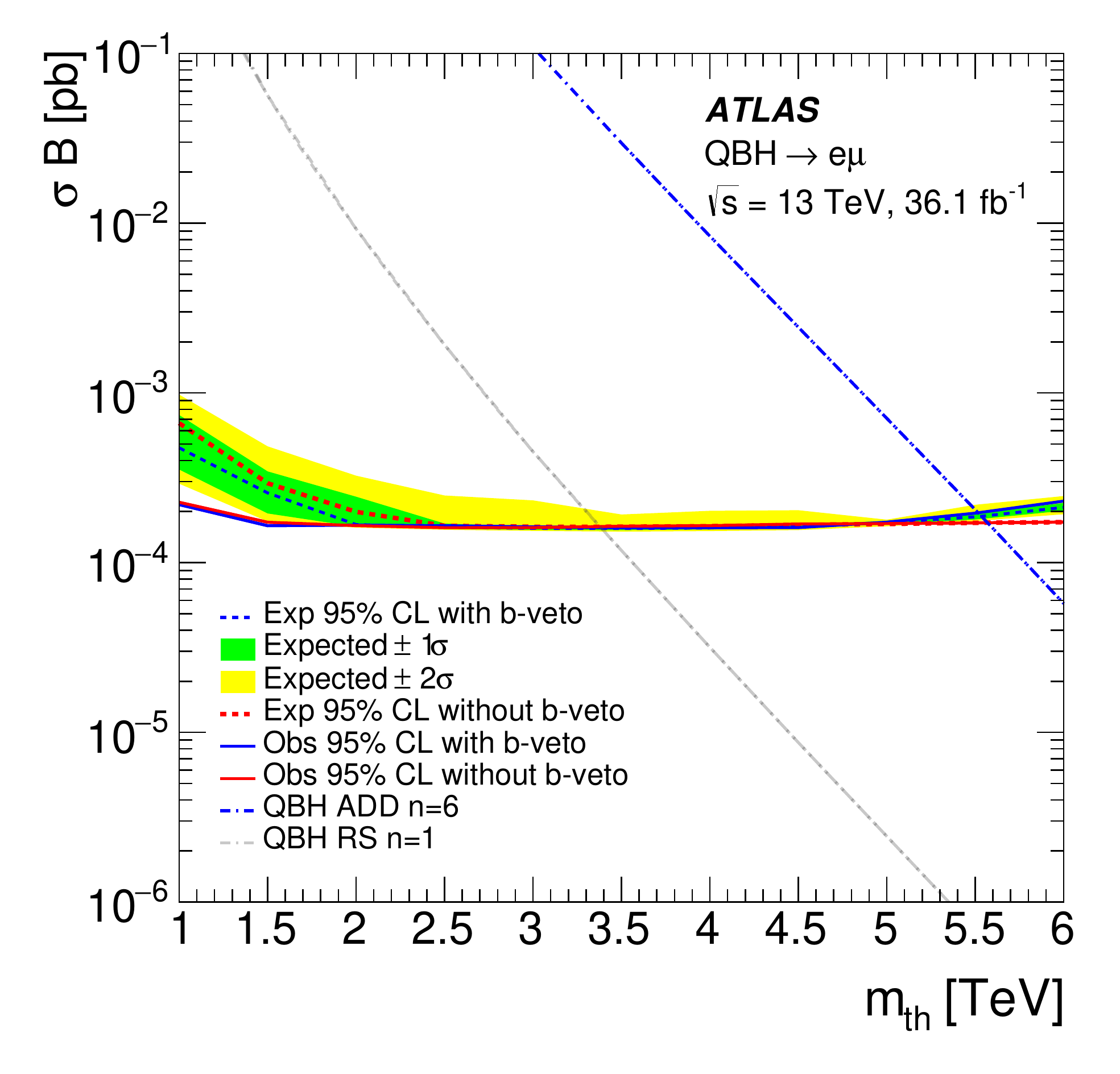}
  }
  \caption{The observed and expected 95\% credibility-level upper limits on the
    (a) \Zprime~boson, (b) $\tau$-sneutrino ($\tilde{\nu}_{\tau}$), and (c) QBH
    ADD and RS production cross-section times branching ratio for decays into an
    $e\mu$ final state with and without the $b$-veto requirement. The signal
    theoretical cross-section times branching ratio lines for the \Zprime~model,
    the QBH ADD model assuming six extra dimensions, and the RS model with one
    extra dimension are obtained from the simulation of each
    process, while the RPV SUSY $\tilde{\nu}_{\tau}$ includes the NLO $K$-factor
    calculated using LoopTools \cite{bib:LoopTools}. The acceptance times
    efficiency of the ADD and RS QBH models agree within 1\%, and the same
    curve is used for limit extraction. The expected limits are shown with the
    $\pm 1$ and $\pm 2$ standard deviation uncertainty bands on the results
    with the $b$-veto requirement.}
  \label{fig:limit_emu}
\end{figure}

\begin{figure}[!htbp]
  \centering
  \subfloat[]{
    \includegraphics[width=0.49\linewidth]{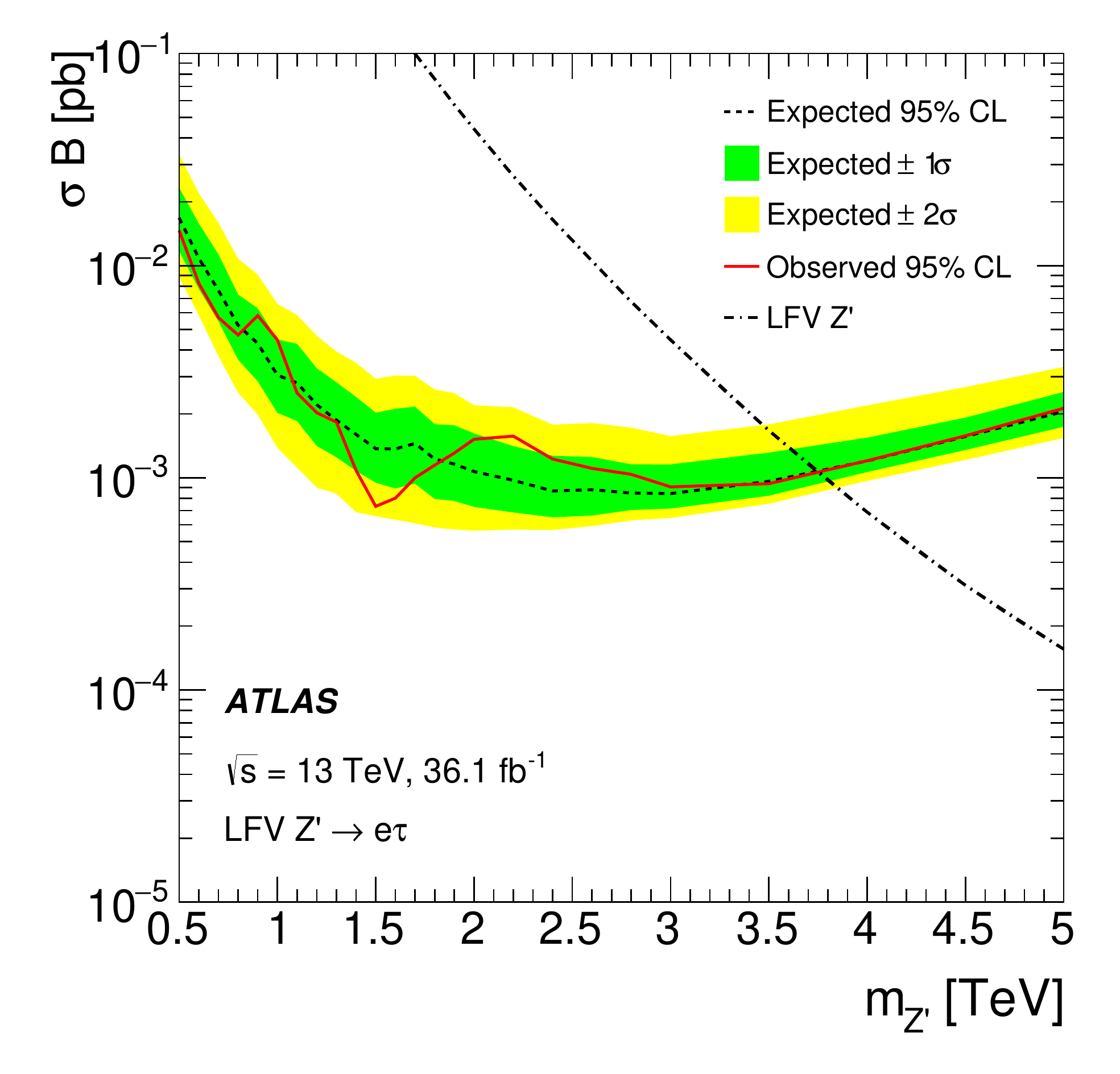}
  }
  \subfloat[]{
    \includegraphics[width=0.49\linewidth]{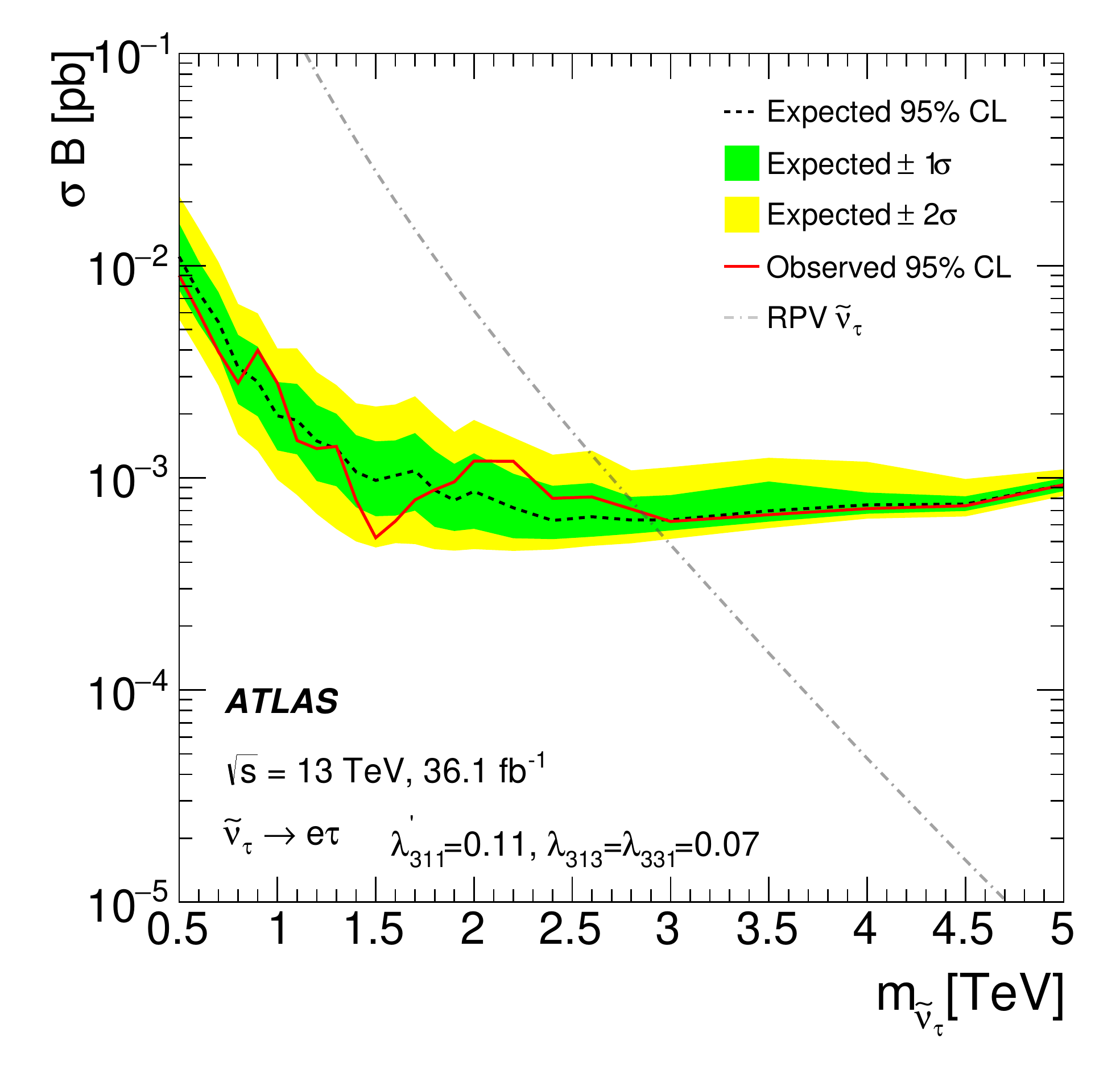}
  } \\
  \subfloat[]{
    \includegraphics[width=0.49\linewidth]{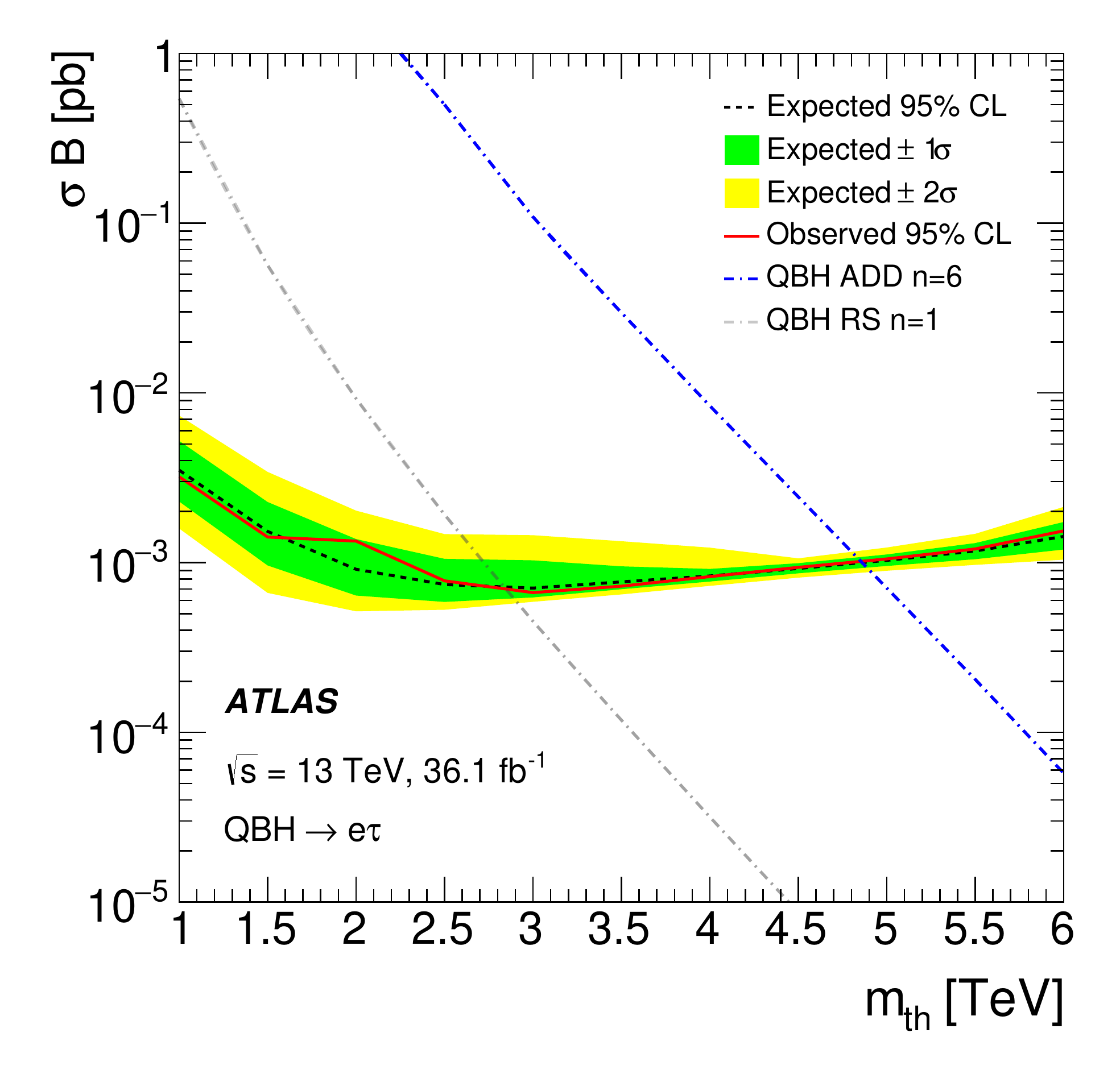}
  }
  \caption{The observed and expected 95\% credibility-level upper limits on the
    (a) \Zprime~boson, (b) $\tau$-sneutrino ($\tilde{\nu}_{\tau}$), and (c) QBH
    ADD and RS production cross-section times branching ratio for decays into an
    $e\tau$ final state. The signal theoretical cross-section times branching
    ratio lines for the \Zprime~model, the QBH ADD model assuming six extra
    dimensions, and the RS model with one extra dimension are obtained from the
    simulation of each process, while the RPV SUSY
    $\tilde{\nu}_{\tau}$ includes the NLO $K$-factor calculated using LoopTools
    \cite{bib:LoopTools}. The acceptance times efficiency of the ADD and RS QBH
    models agree within 1\%, and the same curve is used for limit
    extraction. The expected limits are shown with the $\pm 1$ and $\pm 2$
    standard deviation uncertainty bands.}
  \label{fig:limit_etau}
\end{figure}

\begin{figure}[!htbp]
  \centering
  \subfloat[]{
    \includegraphics[width=0.49\linewidth]{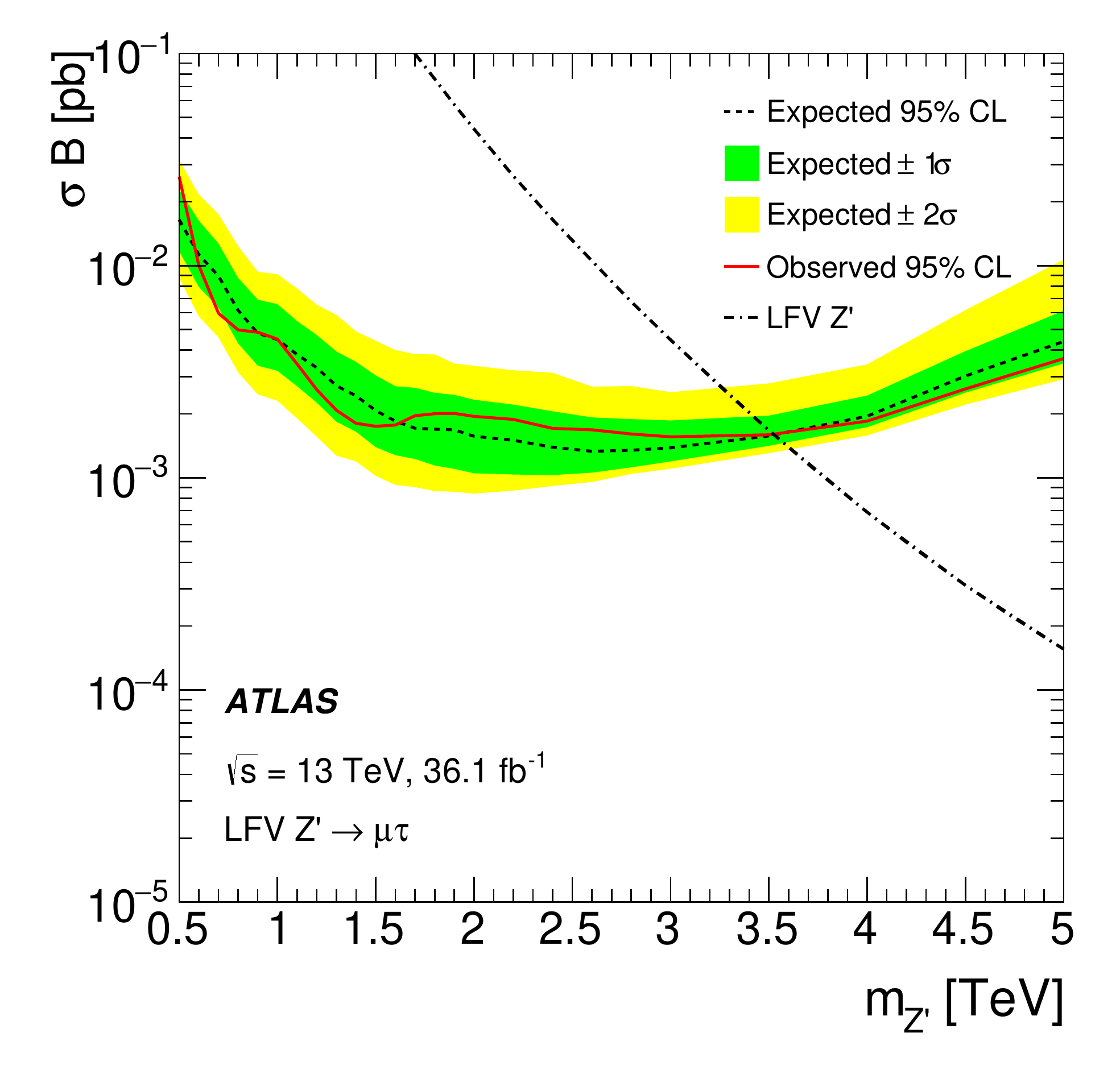}
  }
  \subfloat[]{
    \includegraphics[width=0.49\linewidth]{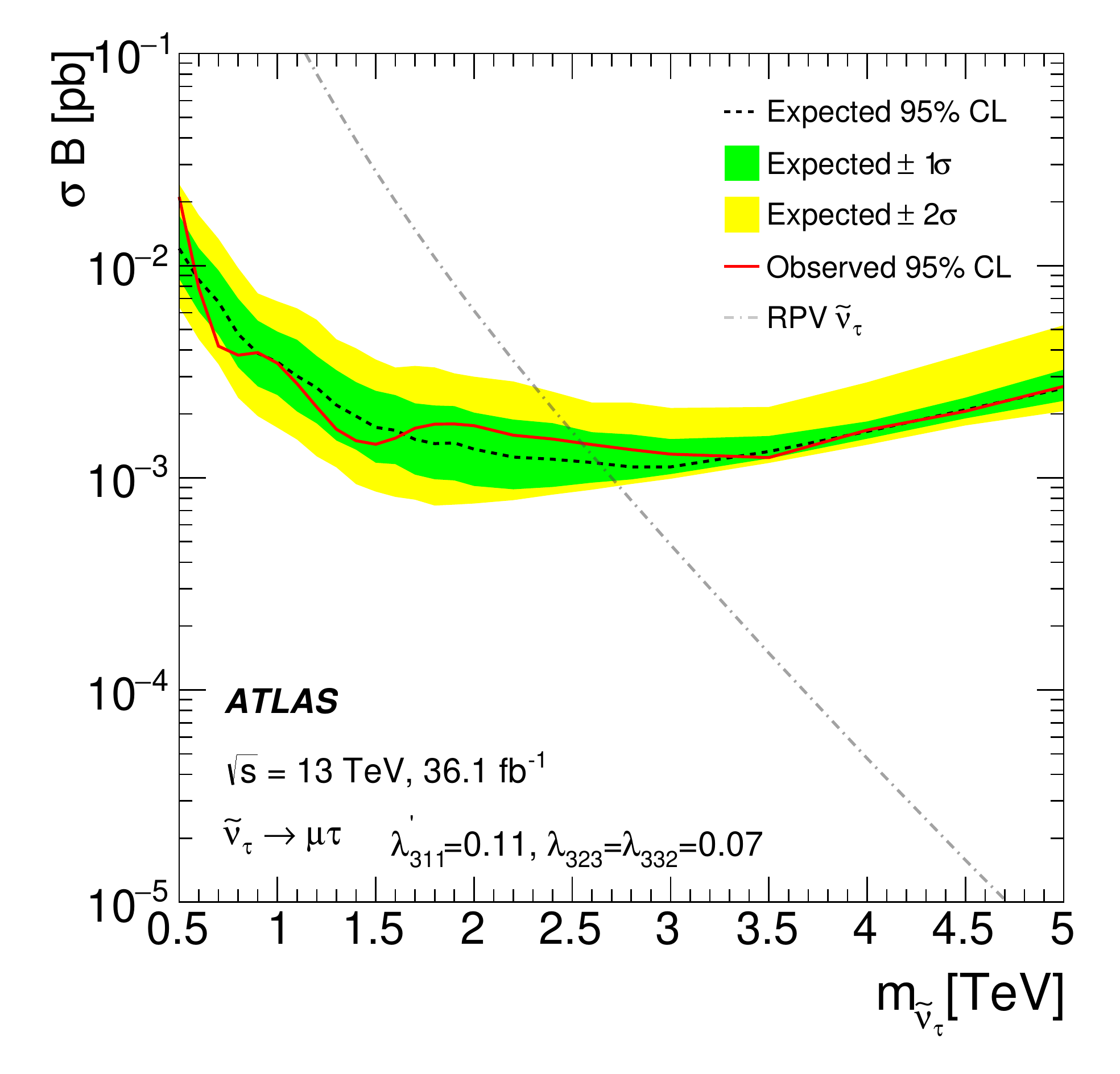}
  } \\
  \subfloat[]{
    \includegraphics[width=0.49\linewidth]{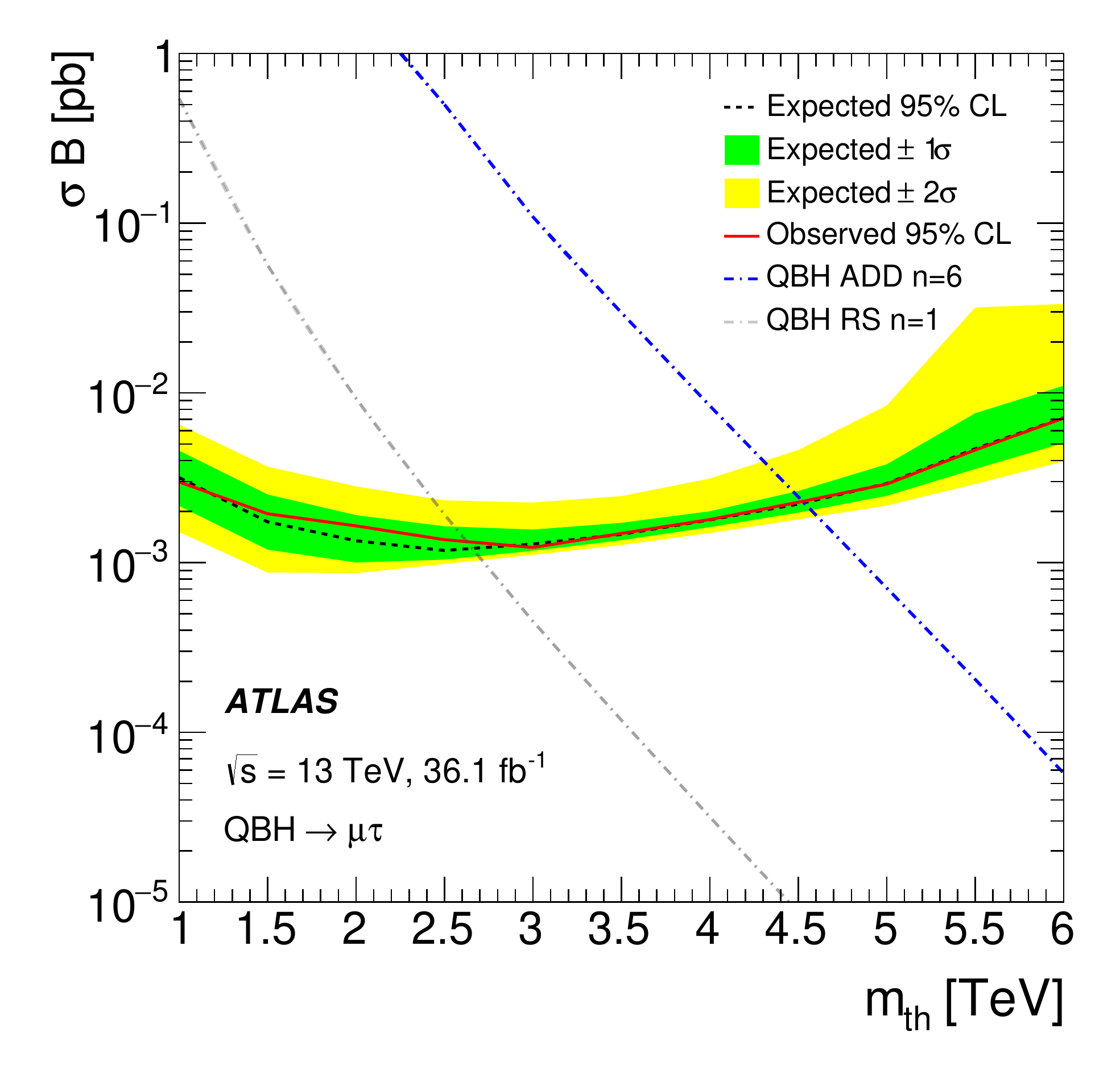}
  }
  \caption{The observed and expected 95\% credibility-level upper limits on the
    (a) \Zprime~boson, (b) $\tau$-sneutrino ($\tilde{\nu}_{\tau}$), and (c) QBH
    ADD and RS production cross-section times branching ratio for decays into an
    $\mu\tau$ final state. The signal theoretical cross-section times branching
    ratio lines for the \Zprime~model, the QBH ADD model assuming six extra
    dimensions, and the RS model with one extra dimension are obtained from the
    simulation of each process, while the RPV SUSY
    $\tilde{\nu}_{\tau}$ includes the NLO $K$-factor calculated using LoopTools
    \cite{bib:LoopTools}. The acceptance times efficiency of the ADD and RS QBH
    models agree within 1\%, and the same curve is used for limit
    extraction. The expected limits are shown with the $\pm 1$ and $\pm 2$
    standard deviation uncertainty bands.}
  \label{fig:limit_mutau}
\end{figure}

Figures~\ref{fig:limit_emu}--\ref{fig:limit_mutau}
show the observed and expected 95\% CL upper limits on the production
cross-section times branching ratio of the \Zprime, RPV SUSY $\tilde{\nu}_{\tau}$
and QBH models for each of the final states considered. The extracted limits
are not as strong for signal masses above about \SI{2.5}{\TeV} due to a
decrease in acceptance at very high $p_{\mathrm{T}}$ and, specifically to the LFV
\Zprime~model, low-mass signal production due to PDF suppression. The results
are summarized in Table~\ref{tab:limits}. The acceptance times efficiency of
the ADD and RS QBH models agree within 1\%, and the same prediction is used for
the limit extraction.
\begin{table}[!htbp]
  \centering
  \caption{Expected and observed 95\% credibility-level lower limits on the
    mass of a \Zprime~boson with lepton-flavor-violating couplings, a
    supersymmetric $\tau$-sneutrino ($\tilde{\nu}_{\tau}$) with
    $R$-parity-violating couplings, and the threshold mass for quantum
    black-hole production for the ADD $n=6$ and RS $n=1$ models.} 
  \begin{tabular}{l|cccc|cccc}
    \hline\hline
    \multirow{3}{*}{Model}
    & \multicolumn{4}{c|}{Expected limit [\si{\TeV}]}
    & \multicolumn{4}{c}{Observed limit [\si{\TeV}]} \\
    \cline{2-9}
    & $e\mu$ &  $e\mu$ & $e\tau$ & $\mu\tau$
    & $e\mu$ &  $e\mu$ & $e\tau$ & $\mu\tau$ \\
    &        &  {\footnotesize ($b$-veto)} &  & 
    &        &  {\footnotesize ($b$-veto)} &  & \\
    \hline
    LFV \Zprime
    & 4.3 &  4.3 & 3.7 & 3.5 & 4.5 & 4.4 & 3.7 & 3.5 \\
    RPV SUSY $\tilde{\nu}_{\tau}$
    & 3.4 &  3.4 & 2.9 & 2.6 & 3.4 & 3.4 & 2.9 & 2.6 \\
    QBH ADD $n=6$
    & 5.6 &  5.5 & 4.9 & 4.5 & 5.6 & 5.5 & 4.9 & 4.5 \\
    QBH RS $n=1$
    & 3.3 &  3.4 & 2.8 & 2.7 & 3.4 & 3.4 & 2.9 & 2.6 \\
    \hline\hline
  \end{tabular}
  \label{tab:limits}
\end{table}

\begin{figure}[!htbp]
  \centering
  \subfloat[]{
    \includegraphics[width=0.49\linewidth]{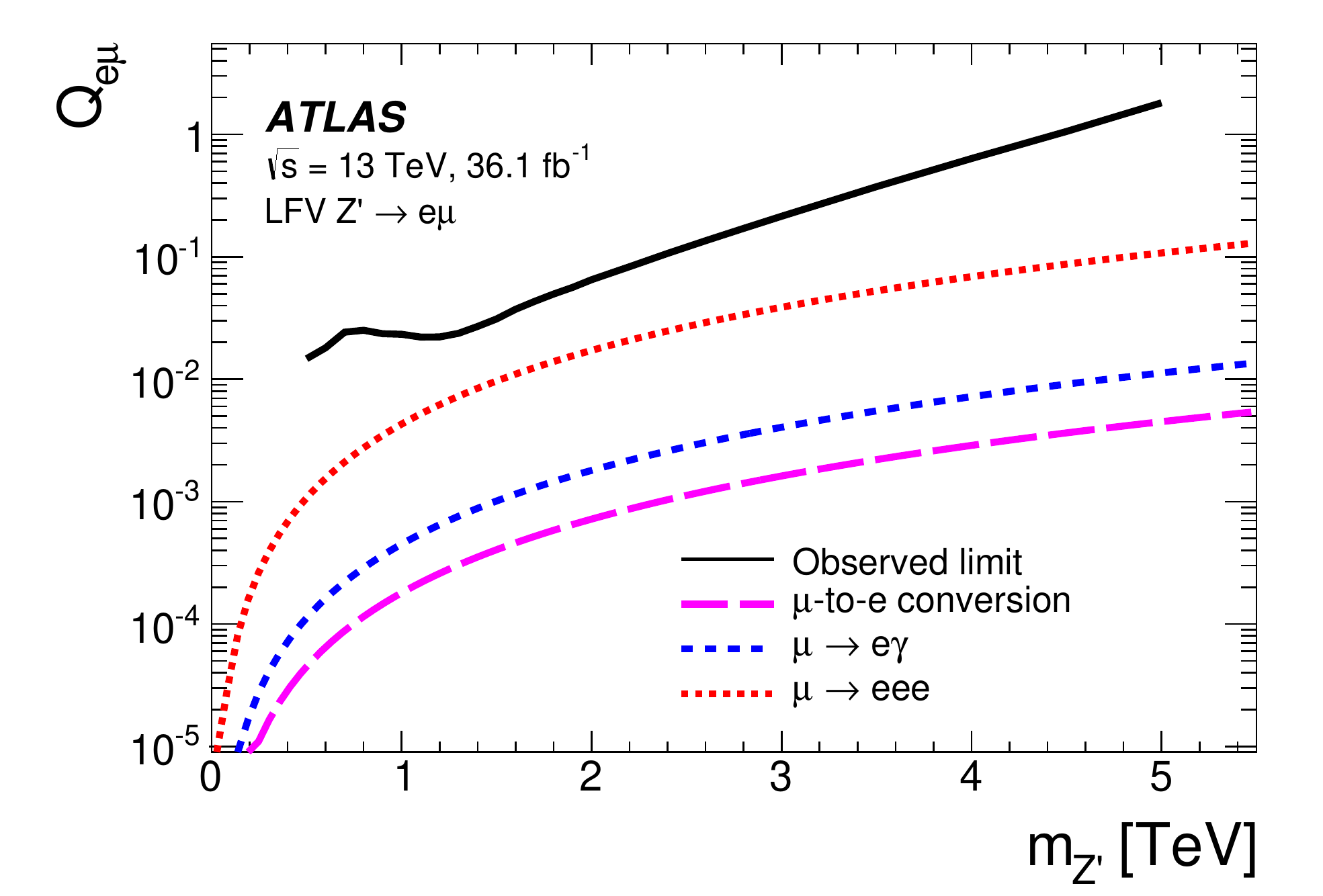}
  }
  \subfloat[]{
    \includegraphics[width=0.49\linewidth]{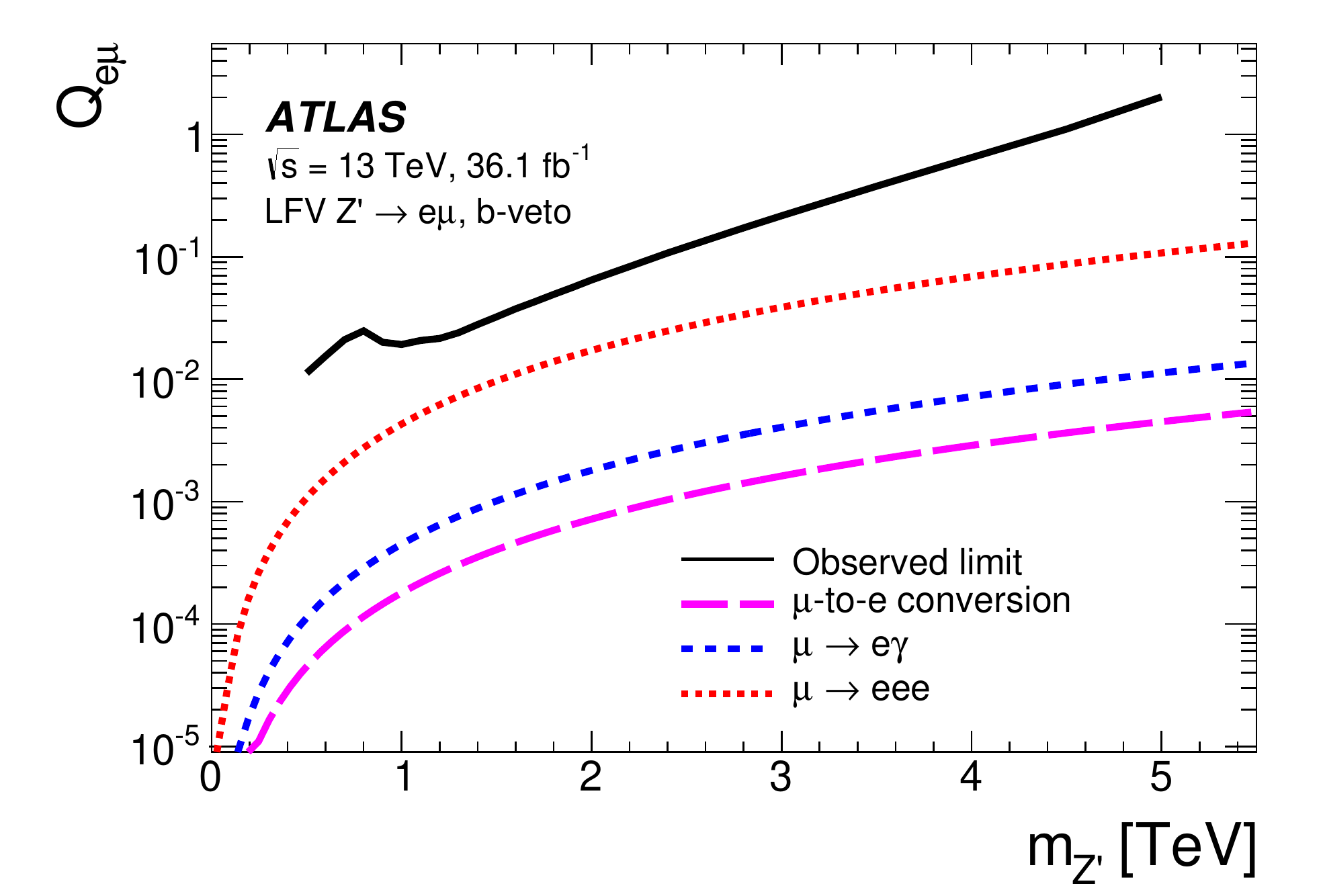}
  }  \\
  \subfloat[]{
    \includegraphics[width=0.49\linewidth]{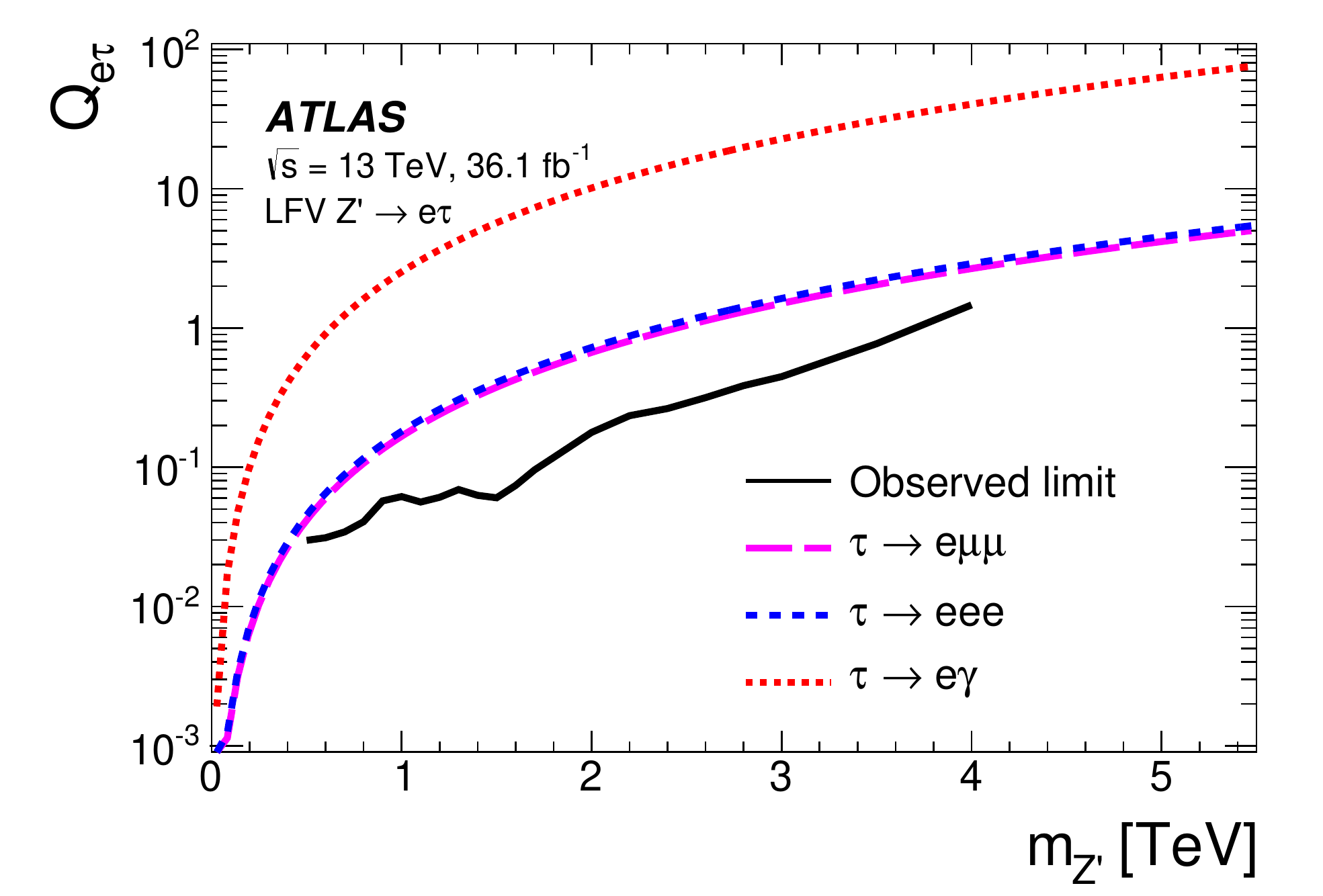}
  } 
  \subfloat[]{
    \includegraphics[width=0.49\linewidth]{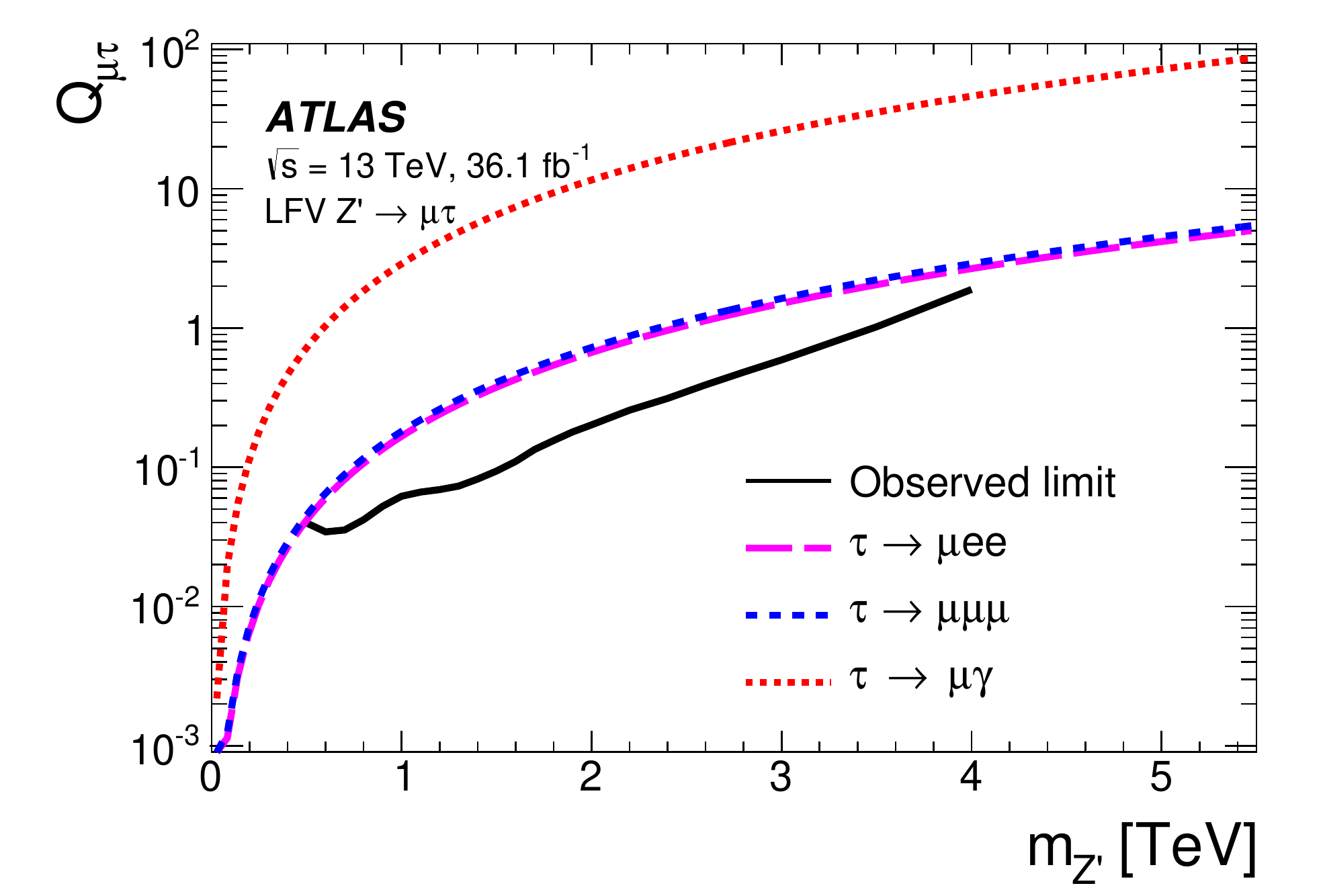}
  } 
  \caption{The 95\% credibility-level upper limits on the couplings
    (a) $Q_{e\mu}$, (b) $Q_{e\mu}$ with a $b$-veto, (c) $Q_{e\tau}$, and
    (d) $Q_{\mu\tau}$ as a function of $m_{\Zprime}$ from the cross-section times
    branching ratio limits in this paper (solid lines) and from low-energy
    experiments (dashed lines).}
  \label{fig:Qlim}
\end{figure}

Results expressed in terms of the coupling limts can be directly compared to
those obtained from precision low energy experiments~\cite{bib:PDG}.
For the \Zprime~model the cross-section times branching ratio is proportional
to $Q^2_{\ell\ell^\prime}$, and the same quark couplings as the SM $Z$ boson are
used. The limits on $Q_{\ell\ell^\prime}$ are shown in Figure~\ref{fig:Qlim} as a
function of $m_{\Zprime}$ for the three channels. The most stringent coupling
limits from low-energy experiments are from $\mu$-to-$e$ conversion and
$\mu \to eee$ for the $e\mu$ channel, from $\tau \to eee$ and
$\tau \to e \mu\mu$ for the $e\tau$ channel, and from $\tau \to \mu\mu\mu$ and
$\tau\to e \mu\mu$ for the $\mu\tau$ channel. The current experimental limits
on these processes are converted to coupling limits using the formulae of
Ref.~\cite{bib:LangPlum2000} and are shown in Figure~\ref{fig:Qlim}. For the
$e\tau$ and $\mu\tau$ channels, the observed limit is restricted up to the
\Zprime~mass point of \SI{4}{\TeV}. This because for the higher mass points,
the limit on $Q_{\ell\ell^\prime}$ becomes sufficiently large that the total width
of the \Zprime~would be significantly larger than the experimental resolution
and violate our assumptions on that. For the $e\mu$ channel, the coupling
limits in this paper do not compete with those from low-energy experiments,
but for the $e\tau$ and $\mu\tau$ channels, the coupling limits in this paper
are more stringent, though they require additional assumptions on the quark
couplings.

\begin{figure}[!htbp]
  \centering
  \subfloat[]{
    \includegraphics[width=0.49\linewidth]{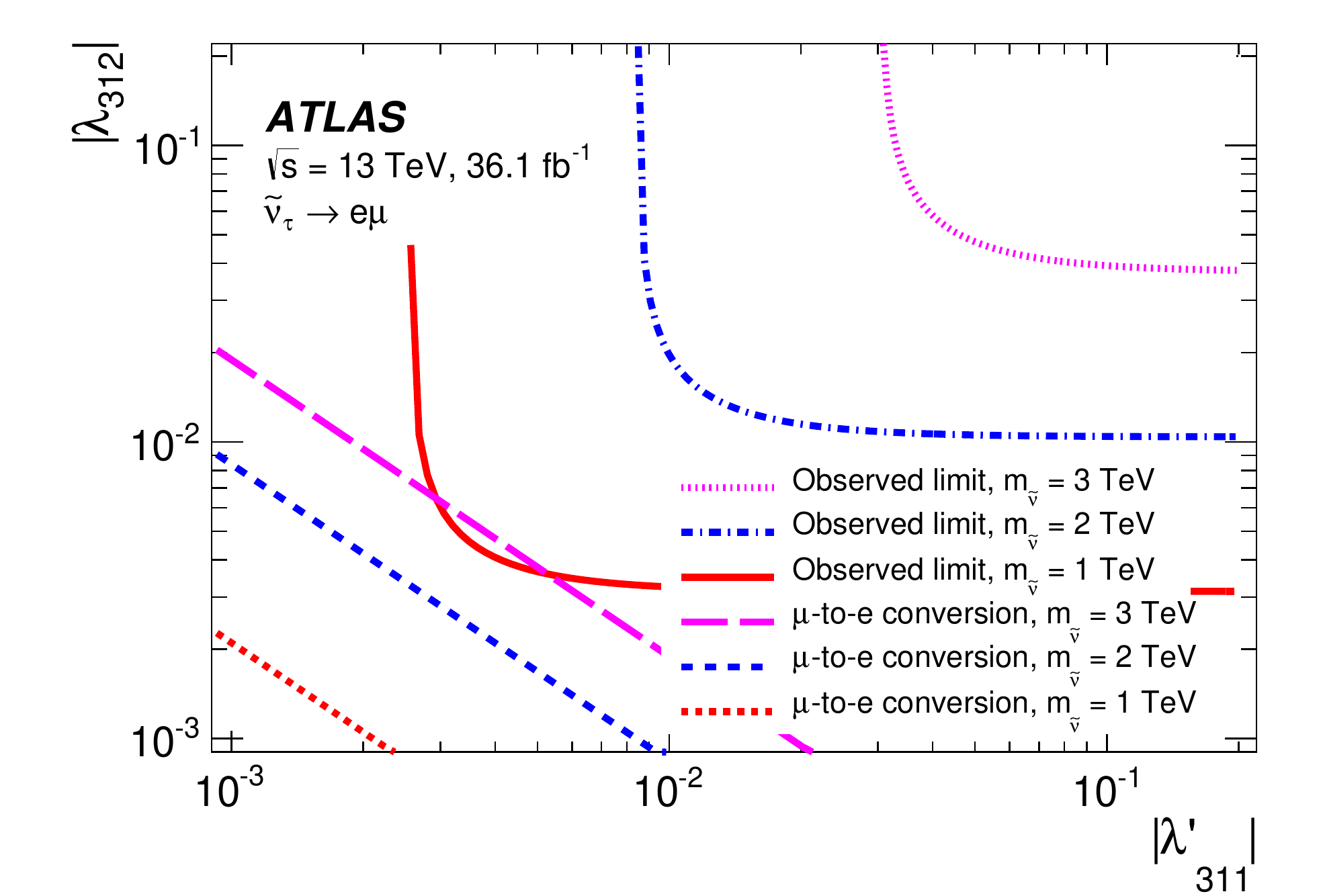}
  }
  \subfloat[]{
    \includegraphics[width=0.49\linewidth]{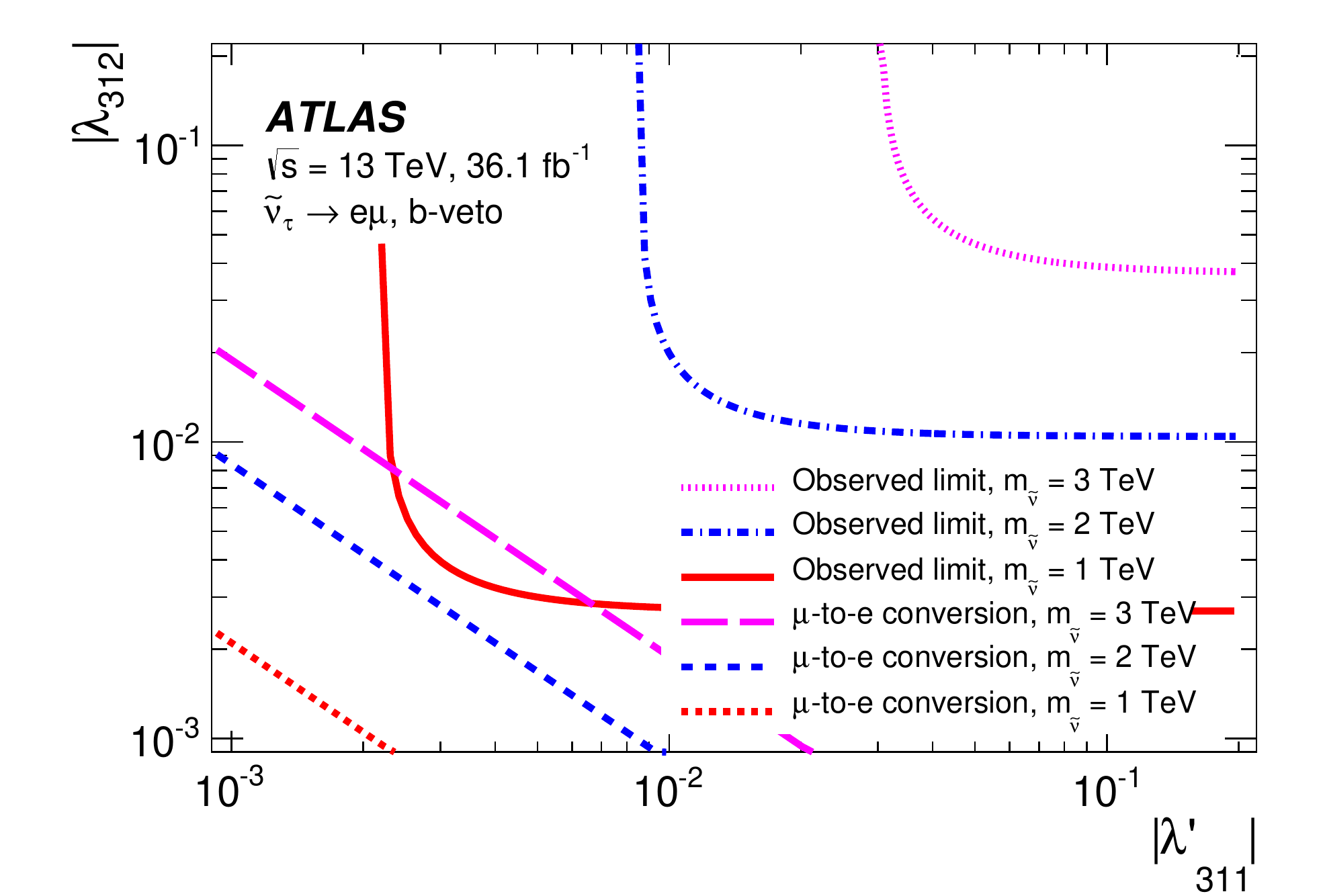}
  }  \\
  \subfloat[]{
    \includegraphics[width=0.49\linewidth]{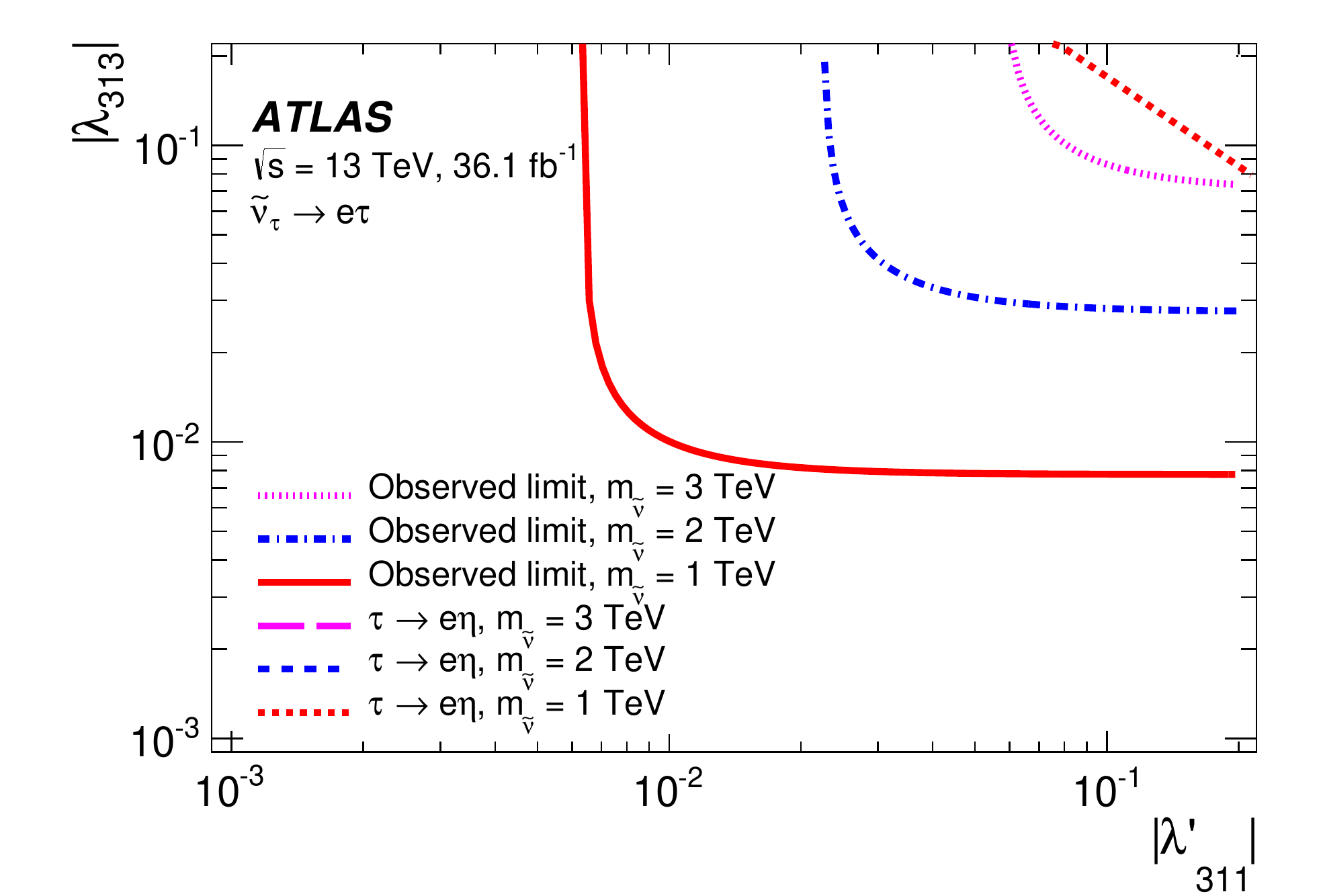}
  }
  \subfloat[]{
    \includegraphics[width=0.49\linewidth]{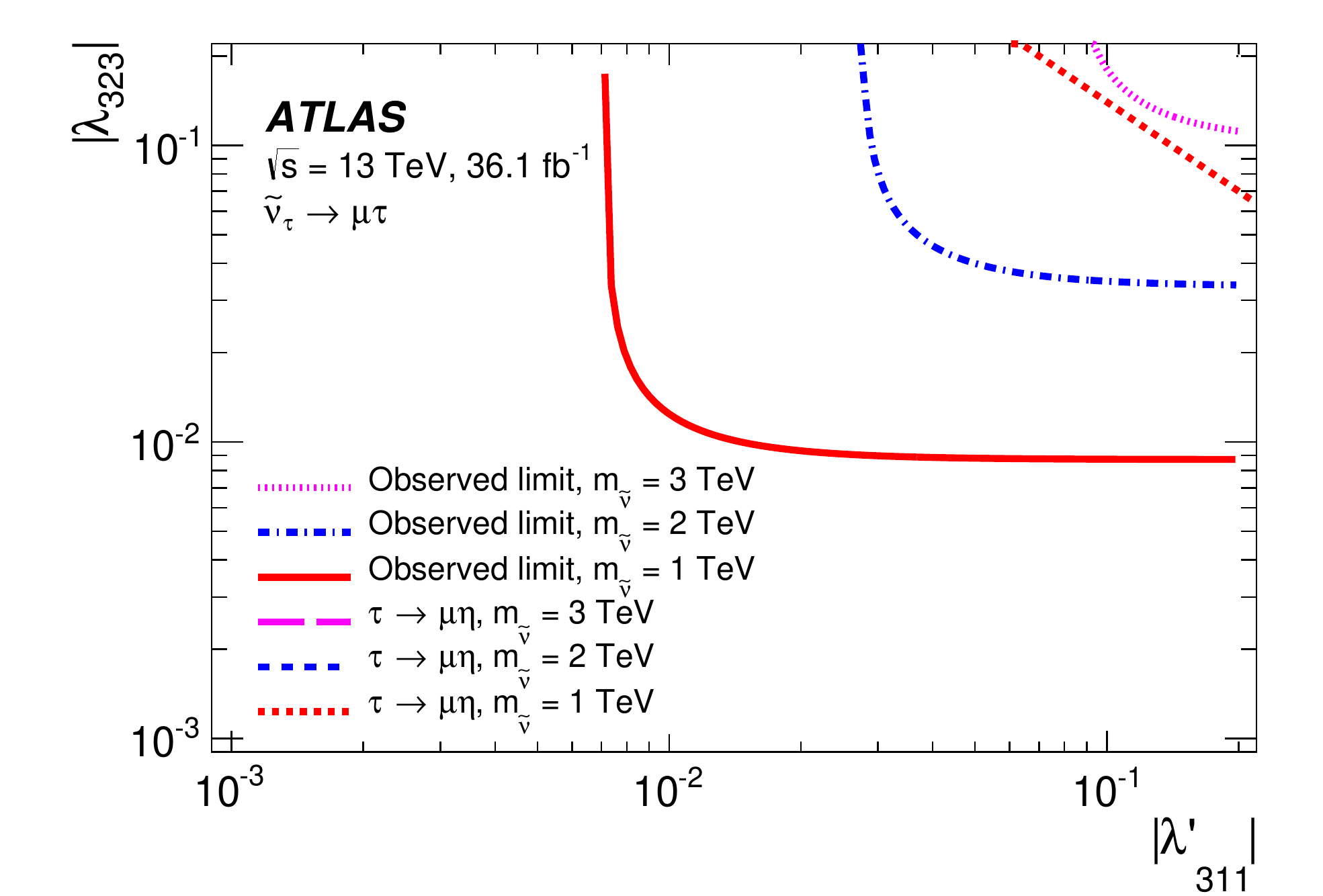}
  } 
  \caption{The 95\% credibility-level upper limits on the RPV couplings
    (a) $|\lambda_{312}|$, (b) $|\lambda_{312}|$ with a $b$-veto, 
    (c) $|\lambda_{313}|$, and (d) $|\lambda_{323}|$ versus
    $|\lambda^\prime_{311}|$ for a few values of $m_{\tilde{\nu}}$ from the
    cross-section times branching ratio limits in this paper (solid, dot-dashed,
    and  dotted lines) and from low-energy experiments (short-dashed,
    medium-dashed, and long-dashed lines).  For the $e\tau$ and $\mu\tau$
    channels, the low energy limits for the 2~\si{\TeV} and 3~\si{\TeV} mass
    points are outside the frame, beyond the upper-right corner.}
  \label{fig:nulim}
\end{figure}

For the $\tilde{\nu}_\tau$ model, the dependence on the couplings is more
complicated because both the production and the decay violate lepton-flavor
conservation. Assuming only the $d\bar{d}$ and $\ell\ell^\prime$ couplings, the
cross-section times branching ratios are proportional to the Yukawa couplings
$|\lambda^\prime_{311}\lambda_{3ij}|^2/(3|\lambda^\prime_{311}|^2+2|\lambda_{3ij}|^2)$,
where $ij$ = 12, 13, and 23 for the $e\mu$, $e\tau$, and $\mu\tau$ channels,
respectively. The factor 3 in the denominator accounts for color and the factor
2 is because both final-state charge combinations are allowed
($\ell^\pm \ell^{\prime \mp}$). The limits on $|\lambda^\prime_{3ij}|$ versus
$|\lambda_{311}|$ are shown in Figure~\ref{fig:nulim} for $\tilde{\nu}_{\tau}$
masses of \SI{1}{\TeV}, \SI{2}{\TeV}, and \SI{3}{\TeV}. 
The most stringent coupling limits set by low-energy experiments derive from
$\mu$-to-$e$ conversion for the $e\mu$ channel, from $\tau \to e\eta$ for the
$e\tau$ channel, and from $\tau \to \mu\eta$ for the $\mu\tau$ channel. The
coupling limits in Ref.~\cite{bib:RParitySUSY} are scaled to current
experimental limits on these processes \cite{bib:Belle07} and are shown in
Figure~\ref{fig:nulim}. For the $e\mu$ channel, the coupling limits in this
paper do not compete with those from low-energy experiments, but for
$e\tau$ and $\mu\tau$ channels, the coupling limits in this paper are more
stringent.

\FloatBarrier

\section{Conclusions}
\label{sec:conclusion}
A search for a heavy particle decaying into an $e\mu$, $e\tau$, or $\mu\tau$
final state is conducted using \SI{36.1}{\per\femto\barn}
of proton--proton collision data at $\sqrt{s}=13$~\TeV\ recorded by the
ATLAS detector at the Large Hadron Collider. The Standard Model predictions are
consistent with the data. From the $e\mu$, $e\tau$, and $\mu\tau$ final states, 
Bayesian lower limits at 95\% credibility level are set on the mass of a
\Zprime~vector boson with lepton-flavor-violating couplings at 4.5, 3.7, and
3.5~\si{\TeV}, respectively; on the mass of a supersymmetric $\tau$-sneutrino
with R-parity-violating couplings at 3.4, 2.9, and 2.6~\si{\TeV}; and on the
threshold mass for quantum black-hole production in the context of the
Arkani-Hamed--Dimopoulos--Dvali (Randall--Sundrum) model at 5.6 (3.4), 4.9
(2.9), and 4.5 (2.6)~\si{\TeV}. The quantum black hole limits extracted are
below those extracted in dijet searches, since the branching ratio to dijet is
expected to be much larger than to dilepton. 
Coupling limits for the lepton-flavor-violating \Zprime~boson and $\tilde{\nu}$
models are more stringent than those from low-energy experiments for the
$e\tau$ and $\mu\tau$ modes.
\clearpage

\section*{Acknowledgements}

We thank CERN for the very successful operation of the LHC, as well as the
support staff from our institutions without whom ATLAS could not be
operated efficiently.

We acknowledge the support of ANPCyT, Argentina; YerPhI, Armenia; ARC, Australia; BMWFW and FWF, Austria; ANAS, Azerbaijan; SSTC, Belarus; CNPq and FAPESP, Brazil; NSERC, NRC and CFI, Canada; CERN; CONICYT, Chile; CAS, MOST and NSFC, China; COLCIENCIAS, Colombia; MSMT CR, MPO CR and VSC CR, Czech Republic; DNRF and DNSRC, Denmark; IN2P3-CNRS, CEA-DRF/IRFU, France; SRNSFG, Georgia; BMBF, HGF, and MPG, Germany; GSRT, Greece; RGC, Hong Kong SAR, China; ISF, I-CORE and Benoziyo Center, Israel; INFN, Italy; MEXT and JSPS, Japan; CNRST, Morocco; NWO, Netherlands; RCN, Norway; MNiSW and NCN, Poland; FCT, Portugal; MNE/IFA, Romania; MES of Russia and NRC KI, Russian Federation; JINR; MESTD, Serbia; MSSR, Slovakia; ARRS and MIZ\v{S}, Slovenia; DST/NRF, South Africa; MINECO, Spain; SRC and Wallenberg Foundation, Sweden; SERI, SNSF and Cantons of Bern and Geneva, Switzerland; MOST, Taiwan; TAEK, Turkey; STFC, United Kingdom; DOE and NSF, United States of America. In addition, individual groups and members have received support from BCKDF, the Canada Council, CANARIE, CRC, Compute Canada, FQRNT, and the Ontario Innovation Trust, Canada; EPLANET, ERC, ERDF, FP7, Horizon 2020 and Marie Sk{\l}odowska-Curie Actions, European Union; Investissements d'Avenir Labex and Idex, ANR, R{\'e}gion Auvergne and Fondation Partager le Savoir, France; DFG and AvH Foundation, Germany; Herakleitos, Thales and Aristeia programmes co-financed by EU-ESF and the Greek NSRF; BSF, GIF and Minerva, Israel; BRF, Norway; CERCA Programme Generalitat de Catalunya, Generalitat Valenciana, Spain; the Royal Society and Leverhulme Trust, United Kingdom.

The crucial computing support from all WLCG partners is acknowledged gratefully, in particular from CERN, the ATLAS Tier-1 facilities at TRIUMF (Canada), NDGF (Denmark, Norway, Sweden), CC-IN2P3 (France), KIT/GridKA (Germany), INFN-CNAF (Italy), NL-T1 (Netherlands), PIC (Spain), ASGC (Taiwan), RAL (UK) and BNL (USA), the Tier-2 facilities worldwide and large non-WLCG resource providers. Major contributors of computing resources are listed in Ref.~\cite{ATL-GEN-PUB-2016-002}.

\clearpage
\printbibliography
\clearpage

 
\begin{flushleft}
{\Large The ATLAS Collaboration}

\bigskip

M.~Aaboud$^\textrm{\scriptsize 34d}$,    
G.~Aad$^\textrm{\scriptsize 99}$,    
B.~Abbott$^\textrm{\scriptsize 124}$,    
O.~Abdinov$^\textrm{\scriptsize 13,*}$,    
B.~Abeloos$^\textrm{\scriptsize 128}$,    
D.K.~Abhayasinghe$^\textrm{\scriptsize 91}$,    
S.H.~Abidi$^\textrm{\scriptsize 164}$,    
O.S.~AbouZeid$^\textrm{\scriptsize 143}$,    
N.L.~Abraham$^\textrm{\scriptsize 153}$,    
H.~Abramowicz$^\textrm{\scriptsize 158}$
% ...
\bigskip
\\

% ...
$^{13}$Institute of Physics, Azerbaijan Academy of Sciences, Baku; Azerbaijan.\\
% ...
$^{34}$$^{(a)}$Facult\'e des Sciences Ain Chock, R\'eseau Universitaire de Physique des Hautes Energies - Universit\'e Hassan II, Casablanca;$^{(b)}$Centre National de l'Energie des Sciences Techniques Nucleaires (CNESTEN), Rabat;$^{(c)}$Facult\'e des Sciences Semlalia, Universit\'e Cadi Ayyad, LPHEA-Marrakech;$^{(d)}$Facult\'e des Sciences, Universit\'e Mohamed Premier and LPTPM, Oujda;$^{(e)}$Facult\'e des sciences, Universit\'e Mohammed V, Rabat; Morocco.\\
% ...
$^{91}$Department of Physics, Royal Holloway University of London, Egham; United Kingdom.\\
% ...
$^{99}$CPPM, Aix-Marseille Universit\'e, CNRS/IN2P3, Marseille; France.\\
% ...
$^{124}$Homer L. Dodge Department of Physics and Astronomy, University of Oklahoma, Norman OK; United States of America.\\
% ...
$^{128}$LAL, Universit\'e Paris-Sud, CNRS/IN2P3, Universit\'e Paris-Saclay, Orsay; France.\\
% ...
$^{143}$Santa Cruz Institute for Particle Physics, University of California Santa Cruz, Santa Cruz CA; United States of America.\\
% ...
$^{153}$Department of Physics and Astronomy, University of Sussex, Brighton; United Kingdom.\\
% ...
$^{158}$Raymond and Beverly Sackler School of Physics and Astronomy, Tel Aviv University, Tel Aviv; Israel.\\
% ...
$^{164}$Department of Physics, University of Toronto, Toronto ON; Canada.\\
% ...
$^{a}$ Also at  Department of Physics, University of Malaya, Kuala Lumpur; Malaysia.\\
$^{b}$ Also at Borough of Manhattan Community College, City University of New York, NY; United States of America.\\
$^{c}$ Also at Centre for High Performance Computing, CSIR Campus, Rosebank, Cape Town; South Africa.\\
$^{d}$ Also at CERN, Geneva; Switzerland.\\
$^{e}$ Also at CPPM, Aix-Marseille Universit\'e, CNRS/IN2P3, Marseille; France.\\
% ...
$^{*}$ Deceased

\end{flushleft}


\end{document}